\documentclass[journal]{IEEEtran}
\ifCLASSINFOpdf

\else

\fi
\usepackage{CJK}
\usepackage{algorithm}
\usepackage{pifont}

\usepackage{algorithmic}
\usepackage{amsmath}
\usepackage{amssymb}
\usepackage{psfrag}
\usepackage{graphicx}
\usepackage{epstopdf}
\usepackage{multirow}
\usepackage{stfloats}
\usepackage{cite}
\usepackage{subcaption}
\usepackage{color}
\usepackage[normalem]{ulem}
\usepackage{arydshln}
\usepackage{enumerate}
\usepackage{bbm}\usepackage{verbatim}
\newtheorem{theorem}{\bf{Theorem}}[section]

\newcommand{\bm}[1]{\mbox{\boldmath{$#1$}}}

\begin{document}

 \title{Impact of Channel Aging on Dual-Function Radar-Communication Systems:  Performance Analysis and  Resource Allocation }

\vspace{-2 cm}
\author{Jie Chen, \IEEEmembership{Member, IEEE}, Xianbin Wang, \IEEEmembership{Fellow, IEEE}, and Ying-Chang Liang, \IEEEmembership{Fellow, IEEE}

\thanks{%This work was supported in part by the National Natural Science Foundation of China under Grants 61631005, U1801261, and 61571100.
 J. Chen and X. Wang are with the  Department of Electrical and Computer Engineering, Western University, London, ON N6A 5B9, Canada (e-mails: chenjie.ay@gmail.com, xianbin.wang@uwo.ca).
  Y.-C. Liang is with Institute for Infocomm Research (I$^2$R), Agency for Science, Technology and Research
(A*STAR), Singapore 138632 (e-mail: liangyc@ieee.org).

}

}
 \maketitle%
\begin{abstract}
In conventional dual-function radar-communication (DFRC) systems, the radar and communication channels are routinely estimated at  fixed time intervals based on their worst-case operation scenarios.
Such situation-agnostic repeated estimations cause significant training overhead and dramatically degrade the system performance, especially for applications with dynamic sensing/communication demands and limited radio resources.
In this paper, we leverage the channel aging characteristics to reduce training overhead and to design a situation-dependent channel re-estimation interval optimization-based resource allocation for performance improvement in a multi-target tracking DFRC system.
Specifically, we exploit the channel temporal correlation to predict radar and communication channels for reducing the need of training preamble retransmission.
 Then, we characterize the channel aging effects on the Cramer-Rao lower bounds (CRLBs) for radar tracking performance analysis and achievable rates with maximum ratio transmission (MRT) and zero-forcing
(ZF) transmit beamforming for communication performance analysis.
In particular, the aged CRLBs and achievable rates are derived as closed-form expressions with respect to the channel aging time,  bandwidth, and power.
Based on the analyzed results, we optimize these factors to maximize the  average total aged achievable rate subject to individual target tracking precision demand, communication rate requirement, and other practical constraints.
Since the formulated problem belongs to a non-convex problem, we develop an efficient one-dimensional search based
optimization algorithm to obtain its suboptimal solutions.
Finally, simulation results are presented to validate the correctness of the derived theoretical results and the effectiveness of the proposed allocation scheme.
\end{abstract}
% Note that keywords are not normally used for peerreview papers.


\IEEEpeerreviewmaketitle

\section{Introduction}

Dual-function radar-communication (DFRC) system has emerged as a promising paradigm in the future generation of wireless systems and networks, particularly in 5G beyond and 6G \cite{zheng2019radar,liu2022integrated,sturm2011waveform}. Due to its capability of fulfilling the stringent sensing and transmission demands concurrently, many new use cases, such as augmented/virtual reality, intelligent transportation, and smart manufacturing, can be supported by DFRC.
In general, a DFRC system is a jointly designed coexistence system, which integrates both radar sensing and communication functions into a unified physical platform by enabling the shared use of signal waveforms, hardware, and radio resources. Given its many advantages and promising potentials, the DFRC system has been under intensive investigation in industrial and academic communities.

Due to the perpetual radio resource constraint, efficient radio resource sharing between sensing and communication in DFRC plays a critical role in the system design to achieve desired performance trade-off between concurrent functions \cite{liu2020joint}.
In general, existing research on resource sharing designs in DFRC systems may be classified into three categories, i.e., communication-centric based schemes
\cite{yang2020queue,ouyang2022noma,tian2022adaptive,temiz2021optimized,luo2019optimization,zhou2021performance,wang2020constrained}, sensing-centric based schemes \cite{wang2022noma,ashraf2022detection,cui2013mimo,hua2022integrated,cao2020joint,ni2021multi,xie2017joint}, joint-centric based schemes \cite{yuan2020spatio,kumari2019adaptive,9913311}.
Specifically, for the communication-centric based schemes, the radio resources were allocated to optimize the communication performance metrics, e.g., energy efficiency \cite{yang2020queue}, outage probability \cite{ouyang2022noma}, bit error rate (BER) \cite{tian2022adaptive}, and achievable minimum/sum/secret rates \cite{temiz2021optimized,luo2019optimization,zhou2021performance,wang2020constrained}, while satisfying the desired radar performance demands. As for sensing-centric based schemes, the resource allocation was investigated to optimize the sensing performance metrics, e.g.,
 desired sensing beam-pattern \cite{wang2022noma}, radar detection probability
\cite{ashraf2022detection,cui2013mimo,hua2022integrated}, and estimation accuracy including mean squared error \cite{cao2020joint} and Cramer-Rao lower bound (CRLB) \cite{ni2021multi,xie2017joint}, subject to the communication performance requirements.
For the joint-centric based schemes, the weighted objective of communication and radar was considered, e.g.,  in   \cite{yuan2020spatio}, the weighted mutual information for sensing and communication was optimized considering the impacts of training overhead and channel estimation error; in \cite{kumari2019adaptive}, the weighted sum of radar CRLB and communication distortion minimum mean
square error (MMSE) was optimized by designing an adaptive virtual transmit signal waveform; and in \cite{9913311}, the sum of power consumption for radar and communication is minimized by joint optimizing active and passive beamforming in an IRS-aided DFRC system.

%a linear target motion state model was studied in \cite{zhou1999tracking} and the maximum likelihood (ML) approach was proposed to track the direction-of-arrival of multiple targets. Then,

Furthermore, the above studies \cite{yang2020queue,ouyang2022noma,tian2022adaptive,temiz2021optimized,luo2019optimization,zhou2021performance,wang2020constrained,
wang2022noma,ashraf2022detection,cui2013mimo,hua2022integrated,cao2020joint,ni2021multi,xie2017joint,
yuan2020spatio,kumari2019adaptive,9913311} focus on sensing-communication performance trade-off in a single transmission block scenario, which cannot be applied to target tracking scenarios with multiple transmission blocks.  In fact, target tracking, an essential task, has been intensively studied  during the past decades \cite{yan2015simultaneous,yan2014power,zhang2020power,muns2019beam,yuan2020bayesian,liu2020radar}.
Specifically, for target tracking designs in conventional radar systems \cite{yan2015simultaneous,yan2014power,zhang2020power}, the multibeam resource and power allocation were studied in  \cite{yan2015simultaneous} and \cite{yan2014power}, respectively, to minimize the maximum Bayesian CRLB among multiple targets,
and the joint power and bandwidth allocation was studied in  \cite{zhang2020power} to minimize the weighted posterior CRLB.
As for target tracking designs in DFRC systems \cite{muns2019beam,yuan2020bayesian,liu2020radar}, the radar ranging method was applied to reduce the training overhead of beam alignment in \cite{muns2019beam}, where the cost-benefit trade-off through time allocation between radar and communication modes was analyzed under the IEEE 802.11ad protocol.
Then, the message passing \cite{yuan2020bayesian},  extended Kalman filtering \cite{liu2020radar}, and deep learning \cite{liu2022learning} methods are proposed to estimate/predict the target kinematic parameters, which are further applied to guide resource allocation for communication performance enhancement.
 

However, most studies  in
\cite{yang2020queue,ouyang2022noma,tian2022adaptive,temiz2021optimized,luo2019optimization,zhou2021performance,wang2020constrained,
wang2022noma,ashraf2022detection,cui2013mimo,hua2022integrated,cao2020joint, ni2021multi,xie2017joint,
yuan2020spatio,kumari2019adaptive,9913311,zhou1999tracking,yan2015simultaneous,yan2014power,zhang2020power,muns2019beam,yuan2020bayesian,liu2020radar,liu2022learning} are designed  to routinely estimate communication channel state information (CSI) or radar sensing information (e.g., target mobility/position)  at each fixed time interval.
Such estimation interval is   designed based on the worst-case scenario among all users without considering the specific channel situation and sensing/communication demands, thus causing redundancy and significant training overhead.
In fact, the practical sensing/communication channels are time-correlated due to the target/user motion continuity \cite{chen2018waveform}.
Such correlations can be explored to predict sensing information and communication CSI \cite{baddour2005autoregressive}. This way, the re-estimation time interval could be increased based on the specific demands, thus reducing training overhead and improving resource utilization efficiency.
Since such time correlations become weaker with a longer time interval, the prediction accuracies for sensing information and communication CSI decrease with time, also called channel aging effect.

Investigating channel aging effect and optimizing channel re-estimation time interval to reduce training overhead have been studied in communication systems
\cite{chopra2016throughput,kong2015sum,papazafeiropoulos2016impact,deng2019intermittent,zheng2021impact}. Specifically,
for single-user multi-input multi-output (MIMO) cases \cite{chopra2016throughput}, the effect of channel aging on the system throughput was characterized and the maximum training time interval was derived based on the minimum throughput condition.
As for multi-user MIMO cases \cite{kong2015sum,papazafeiropoulos2016impact,deng2019intermittent},  the sum rate with the aged CSI was evaluated in \cite{kong2015sum,papazafeiropoulos2016impact} with considering the time allocation trade-off between training overhead and data transmission. In \cite{deng2019intermittent},  an intermittent channel
estimation scheme was proposed to reduce the training overhead, and the total throughput was maximized by optimizing training time interval for each user.
Moreover, for cell-free MIMO cases \cite{zheng2021impact}, the spectral efficiencies (SEs) were derived in uplink and downlink transmissions, and the source block length was designed to mitigate the negative effect of channel aging on SE.  Despite the significant benefits of investigating channel aging in communication systems, to the best of our knowledge, there is no study considering channel aging in DFRC systems, where the radar-communication performance analysis and resource allocation design is still an open problem.

Motivated by the above reasons, in this paper, we characterize the channel aging effects on the radar-communication performances and apply them to design a situation-dependent resource allocation scheme for a downlink multi-target tracking DFRC system, in which a full-duplex base station (BS) concurrently communicates with multiple communication users and tracks multiple targets. This system is operated with a novel frame structure consisting of multiple fading blocks to support radar target tracking, communication channel estimation, and information transmission. Besides, the mobility information of targets and CSI of communication users are estimated in the first block by transmitting training preambles, then predicted in the remaining blocks by leveraging channel aging characteristics without repeated training preamble transmission. Then, resource utilization efficiency can be improved by designing a resource allocation scheme and optimizing channel re-estimation interval with considering the specific radar and communication demands. Finally, to highlight the contributions of this paper, we summarize the paper as follows:
\begin{itemize}
\item  To the best of our knowledge, it is the first time to characterize the channel aging effects for the DFRC system. Specifically, we characterize the channel aging effects on the tracking performance (i.e., the CRLBs of the predicated mobility information) and the communication performance (i.e., the achievable rates with the predicated CSI under maximum ratio transmission (MRT) and zero-forcing (ZF) transmit beamforming).
In particular, the aged CRLBs and achievable rates are derived as closed-form expressions with respect to the allocated bandwidth and power resource and channel aging time. Moreover, to intuitively understand channel aging effects, we propose an approximation method to simplify the derived CRLB expressions and provide the asymptotic analysis of the achievable rate with respect to the transmit power.

\item Based on the derived aged tracking and communication performances,
a situation-dependent joint channel re-estimation interval together with other radio resource allocation scheme is proposed for system training overhead reduction and performance improvement. Particularly, we formulate an average total aged achievable rate maximization problem subject to individual tracking precision demand, customized communication rate requirement, and other practical constraints.
Since the formulated problem belongs to a non-convex mixed integer nonlinear programming (MINLP) problem,  an  efficient one-dimensional search based optimization algorithm is developed to obtain its suboptimal solutions.

\item Simulation results show that the simulated CRLBs and achievable rates can approach the derived theoretical results. Besides, the performance of the proposed algorithm is close to the upper bound performance and is significantly superior to other benchmarks.
\end{itemize}

The rest of this paper is organized as follows.
Section II presents the multi-target/-user DFRC system model.
Section III introduces radar and communication channel estimation algorithms and then characterizes the channel aging effects on the tracking and communication performances. Then, Section IV exploits the characterized results to study the average total aged achievable rate maximization problem and provides the
corresponding efficient solutions. Finally, Section V provides simulation results and Section VI concludes the paper.

{\emph {Notation}}:  We use $\mathbb N$  and ${\bf I}_M$ to denote the set of natural numbers and the {\mbox{$M$-by-$M$}} identity matrix, respectively.
We use
${\mathbb E}(\cdot)$ and ${\rm var}(\cdot)$ to denote the expectation and variance of the variable, respectively, and use  {${\rm diag}(\cdot)$ to return a square diagonal matrix with the corresponding elements of a vector or a column vector of the main diagonal elements of a matrix.
Finally,  the {circularly symmetric complex Gaussian} (CSCG) distribution with mean $\mu$ and variance $ {\sigma ^2}$ is denoted as ${\cal C}{\cal N}\left( {\mu,{\sigma ^2}} \right)$.

\section{System Model  }\label{sec:System Model}

\begin{figure}
    \centering
     \includegraphics[width =  .5\textwidth ]{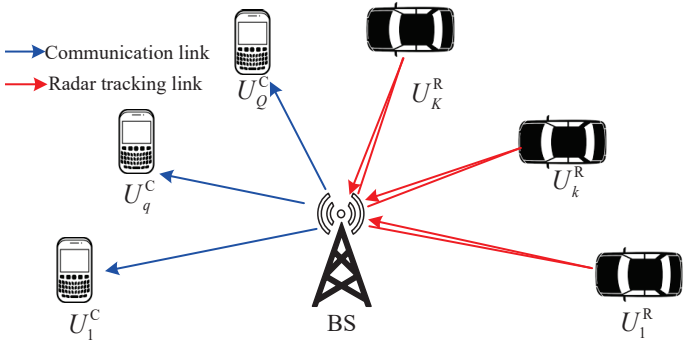}
 \caption{  A downlink DFRC system with one full-duplex BS, $K$ tracking targets, and $Q$ communication receivers. }\label{figp1}
\end{figure}

%
% \begin{figure}
%  \begin{minipage}{.43 \textwidth}
%   \centering
%  \includegraphics[width =  \textwidth, height= 0.55 \textwidth]{SystemModel.eps}
% \caption{  A downlink DFRC system with one full-duplex BS, $K$ tracking targets, and $Q$ communication receivers. }\label{figp1}
%  \end{minipage} \quad
% \begin{minipage}{.45 \textwidth}
%   \centering
%  \includegraphics[width = \textwidth]{FrameStructure.eps}
%\caption{ Transmission frame structure, where each frame includes $N$ blocks and each block is with $M$ symbol durations. } \label{figp2}
%  \end{minipage}
% \end{figure}
As shown in Fig.~\ref{figp1}, we consider a downlink DFRC system, which is composed of one full-duplex BS equipped with $L_t$ transmit
antennas and $L_r$ receive antennas, $K$ radar tracking targets, denoted by ${ U}_k^{\rm R}$  for $1 \le k \le K$, and $ Q$ single-antenna
communication receivers, denoted by ${ U}_q^{\rm C}$ for $ 1\le q \le Q$. In order to reduce the training overhead and achieve high-efficient dual functions of mobility tracking for radar targets and information transmission for communication receivers, we need to first characterize the channel aging effects on the radar-communication performances with respect to the transmission time, and then propose the resource allocation scheme to dynamically optimize the channel re-estimation interval based on the system performance constraints. Hence, we propose  the transmission frame structure consisting of multiple frames, as shown in Fig.~\ref{figp2}.
Specifically,  each frame includes $N$ transmission blocks and each block is with $M$ symbol durations.
In particular, the first block of each frame is divided into two phases, i.e., Phase-I for target tracking and channel estimation and Phase-II for information transmission.
In Phase-I with $M_1$ symbol durations, the BS transmits training signals to track the mobility information of all targets. Simultaneously, the communication receivers utilize the received training signals to perform channel estimation and send feedback to the BS\footnote{Note that there will exist delay and quantification errors during the feedback processing. Low-resolution quantized CSI increases channel estimation error while higher-resolution quantized CSI  increases feedback overhead. Both of them will degrade the transmission performance. Therefore, there is a performance trade-off between the quantification error and CSI feedback overhead. Considering the limited pages and system design complexity, we assume that the feedback delay and quantification error can be ignored and leave the relevant exploration of the influences of these factors as future extension works.}.
In Phase-II with the remaining ($M-M_1$) symbol durations, the BS transmits independent information to communication receivers by utilizing beamforming technology with the estimated CSI in Phase-I.

\begin{figure}
    \centering
   \includegraphics[width = .5\textwidth]{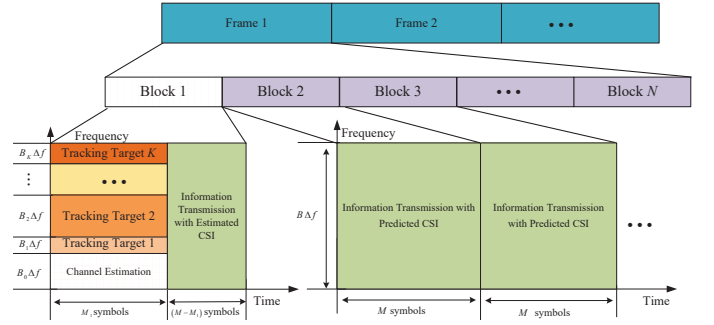}
\caption{ Transmission frame structure, where each frame includes $N$ blocks and each block is with $M$ symbol durations. } \label{figp2}
\end{figure}

Moreover, we assume that the radar and communication channels are time-correlated among all transmission blocks in each frame.
Thus, in the remaining ($N-1$) blocks,  the BS can predict the mobility information of all targets and the CSI of all communication users using the information estimated in the first block.
By adopting this approach, the repeated transmission of dedicated signals for channel estimation and target tracking in the remaining blocks is eliminated, which can lower the training overhead and free up all radio resources for information transmission\footnote{
Although communication signals in the remaining ($N-1$) blocks can be applied to target tracking, as shown in [15], we need to carefully design the downlink beamforming matrix to guarantee transmission and tracking performance. This is because the deployed large-scale antennas at the BS and separated target/receiver locations  make the communication and radar channels nearly independent and orthogonal. Hence, it will not only decrease the communication performance but also increase the design complexity, thus making the performance analysis of the channel aging time effect very challenging. Besides, this paper focuses on channel aging time performance analysis instead of beamforming design, thus we only use the communication signal for data transmission in this paper.}.

However, due to channel aging effects, the accuracies of the predicted mobility information and communication CSI may not support system performance requirements when the time increases. Therefore, we need to characterize the channel aging effects on the radar and communication performances, and then design the resource allocation scheme to improve the re-estimation interval $N$ for overhead reduction and performance improvement subject to the specific sensing and communication demands for resource utilization efficiency improvement.

% \vspace{-0.4cm}
\subsection{Signal Model}

In this part, we introduce the signal models for radar target tracking and communication channel estimation in Phase-I of the first block, and information transmission in Phase-II of the first block and the remaining blocks.

As shown in Fig. \ref{figp2}, in Phase-I of the first block,  the transmitted training signals for channel estimation and radar target tracking from the BS can be given by
%%\begin{align}
%%%{\bf{s}}\left( t \right) = {\sqrt {{p_0}}   {{{\bf{w}}_0}\left( t \right)} {s_0}\left( t \right)} + {\sum\limits_{k = 1}^K {\sqrt {{p_k}} {{\bf{w}}_k}(t){s_k}\left( t \right)} \;},0 \le t \le {M_1}{T_o},\\
%%{\bf{s}}\left( t \right)= \underbrace {{\sqrt {{p_0}}   {{{\bf{w}}_0}\left( t \right)} {s_0}\left( t \right)}}_{{\rm{Channel}}\;{\rm{estimation}}} + \underbrace {\sum\nolimits_{k = 1}^K {\sqrt {{p_k}} {{\bf{w}}_k}(t){s_k}\left( t \right)} \;}_{{\rm{Target}}\;{\rm{tracking}}},0 \le t \le {M_1}{T } ,\label{eqradartraining1}
%%\end{align}
\begin{align}
%{\bf{s}}\left( t \right) = {\sqrt {{p_0}}   {{{\bf{w}}_0}\left( t \right)} {s_0}\left( t \right)} + {\sum\limits_{k = 1}^K {\sqrt {{p_k}} {{\bf{w}}_k}(t){s_k}\left( t \right)} \;},0 \le t \le {M_1}{T_o},\\
{\bf{s}}\left( t \right)&= \underbrace {{\sqrt {{p_0}}   {{{\bf{w}}_0}\left( t \right)} {s_0}\left( t \right)}}_{{\rm{Channel}}\;{\rm{estimation}}} + \underbrace {\sum\nolimits_{k = 1}^K {\sqrt {{p_k}} {{\bf{w}}_k}(t){s_k}\left( t \right)} \;}_{{\rm{Target}}\;{\rm{tracking}}},
\nonumber\\
&\qquad\qquad\qquad\qquad\qquad\qquad\qquad0 \le t \le {M_1}{T } ,\label{eqradartraining1}
\end{align}
where $p_k$ and  ${\bf w}_k(t)\in{\mathbb C}^{L_t\times1}$ denote the downlink transmit power and beamformer with normalized power on the $k$-th band at time instant $t$, respectively.
Here, $s_k\left( t \right)$ is the orthogonal frequency-division multiplexing (OFDM) modulated training signal on the $k$-th frequency band with $B_k$ subcarriers, which is used to estimate the channels of all communication users if $k=0$ or track the $k$-th target if $1\le k \le K$, i.e.,
%\begin{align}{s_k}\left( t \right) = \sum\nolimits_{m = 0}^{{M_1-1}}  \sum\nolimits_{b = 0}^{{B_k-1}}& { \tilde s_{m,b}^k}  {e^{j2\pi \left( {\sum\nolimits_{i =0}^{k - 1} {{B_i}}  + b} \right)\Delta_f\left( {t - {T_{cp}} - m{T }} \right)}}{\rm{rect}}\left( {t - m{T }} \right), 0 \le t \le {M_1}{T },
%\end{align}
\begin{align}
{s_k}\left( t \right) &= \sum\limits_{m = 0}^{{M_1-1}}  \sum\limits_{b = 0}^{{B_k-1}}  {{\tilde s_{m,b}^k}  {e^{j2\pi \left( {\sum\limits_{i =0}^{k - 1} {{B_i}}  + b} \right)\Delta_f\left( {t - {T_{cp}} - m{T }} \right)}} } \nonumber\\
&\qquad\qquad\qquad \times  {{\rm{rect}}\left( {t - m{T }} \right)} , 0 \le t \le {M_1}{T },
\end{align}
where ${\tilde s_{m,b}^k}$ is the complex modulation symbol with power $\frac{1}{B_k}$ transmitted on the $k$-th band $b$-th subcarrier of the $m$-th OFDM symbol, $T_{cp}$ is the duration of cyclic prefix, $T_o=\frac{1}{\Delta_f}$  is the OFDM elementary symbol duration, and $T=T_o+T_{cp}$ is the OFDM symbol duration including the cyclic prefix. Here, $B=\sum\nolimits_{k=0}^KB_k$ and $\Delta_f$ are the total number of subcarriers and bandwidth of each subcarrier, respectively.  The rectangular function is defined as ${\rm{rect}}(x)=1 $ if $0\le x\le T $, otherwise ${\rm{rect}}(x)=0 $.

Next, in Phase-II of the first block and the remaining ($N-1$) blocks, let
$c_{q,n}(t)\sim {\cal C}{\cal N}\left( {0,1} \right)$  be the data symbol transmitted to receiver ${ U}_q^{\rm C}$  at time instant $t$ during the $n$-th block. Then, the transmitted information signals  to all communication receivers can be expressed as
\begin{align}
{\bf{s}}\left( t \right) = &\sum\limits_{q = 1}^Q {\sqrt {{{\tilde p}_{q,n}}} {{\bf{f}}_{q,n}}} {c_{q,n}}\left( t \right), \nonumber\\
&t \in \left\{ {\begin{array}{*{20}{l}}
{\left[ {{M_1}{T },n{\widetilde T }} \right],\; {\rm if }\;n = 1},\\
{\left[ {\left( {n - 1} \right) {\widetilde T },n  {\widetilde T }} \right], \;{\rm if }\;2 \le n \le N},
\end{array}} \right.\label{eq3}
\end{align}
%\begin{align}
%{\bf{s}}\left( t \right) = &\sum\limits_{q = 1}^Q {\sqrt {{{\tilde p}_{q,n}}} {{\bf{f}}_{q,n}}} {c_{q,n}}\left( t \right), t \in \left\{ {\begin{array}{*{20}{l}}
%{\left[ {{M_1}{T },n{\widetilde T }} \right],\; {\rm if }\;n = 1},\\
%{\left[ {\left( {n - 1} \right) {\widetilde T },n  {\widetilde T }} \right], \;{\rm if }\;2 \le n \le N},
%\end{array}} \right.\label{eq3}
%\end{align}
where $\widetilde T=MT$. Here, ${\tilde p}_{q,n}$ and ${\bf f}_{q,n}(t)\in{\mathbb C}^{L_t\times1}$  are the downlink transmit power and beamforming vector with normalized power, respectively, for receiver ${ U}_q^{\rm C}$ at time instant $t$  during the $n$-th block.

\subsection{  Radar and Communication Channel Aging Models}\label{Sec:systemagedmodel}
In this part, we introduce the radar and communication channel aging models, which are the basics to predict the target mobility information and communication CSI.
%mobility information of radar targets and the CSI of communication receivers.

\subsubsection{Radar Channel Aging Model}

In the $n$-th transmission block, we regard the mobility information of target $U_k^{\rm R}$ as the information of its azimuth angle ${\theta _{k,n}}$, distance ${d_{k,n}}$, and velocity ${v_{k,n}}$ relative to the BS\footnote{
Here, we assume that the uniform linear array is deployed at the BS, and we concentrate solely on the azimuth angle of the radar target, disregarding the elevation angle. As for the three-dimensional mobility tracking, it requires the deployment of the uniform planar array, which can be explored in a future extension due to page constraints.}, which is expressed by ${{\bf{x}}_{k,n}} = {\left[ {{\theta _{k,n}},{d_{k,n}},{v_{k,n}}}\right]^T} \in {\mathbb C}^{3\times1}$ .
Based on the  geometric relations between two successive blocks, as shown  in Fig. \ref{fig3}, the target mobility state evolution model can be expressed as \cite{liu2020radar,liu2022learning}
\begin{align}\label{eqRadarState}
\!\!\left\{ \begin{array}{l}
\!\!\!\!d_{k,n}^2\! =\! d_{k,n - 1}^2 \!+\! \Delta_{d_{k,n - 1}}^2 \!\!\!-\! 2{d_{k,n - 1}}\Delta_{d_{k,n - 1}}\!\cos ( \tilde \theta_{k,n-1} ),\\
\!\!\!\!\frac{\Delta_{d_{k,n - 1}}}{{\sin \left( {{\theta _{k,n}} - {\theta _{k,n - 1}}} \right)}}\! = \! \frac{{{d_{k,n}}}}{{\sin \left( \tilde \theta_{k,n-1}\right)}},
\end{array} \right.
\end{align}
where ${\Delta _{{d_{k,n - 1}}}} = {v_{k,n - 1}}\widetilde T$ and $\tilde \theta_{k,n-1}=\theta_{k,n-1}-\varphi_k$. Here,  $ \varphi_k$ is the direction of velocity of $U_k^{\rm R}$  with respect to the negative horizontal direction of the BS. We assume that it  is known at the BS and keeps constant in each block.
\begin{figure}[t]
\centering
\includegraphics[width = 0.45\textwidth]{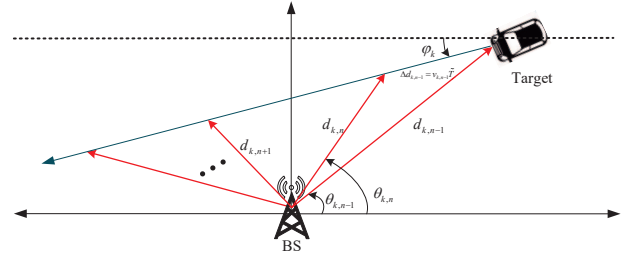}\vspace{-0.1cm}
\caption{Target mobility state evolution model. } \label{fig3}\vspace{-0.4cm}
\end{figure}

By using the similar approximation method in \cite{liu2020radar}, the   state evolution model in \eqref{eqRadarState} can be approximated by
\begin{align}
 \!\!\!\left\{ \begin{array}{l}
\!\!\!{\theta _{k,n}} \!=\! {\theta _{k,n - 1}} + d_{k,n - 1}^{ - 1}{v_{k,n - 1}} \widetilde T\sin ( {\tilde \theta_{k,n-1} } )\! + \!u _{k,n }^\theta, \\
\!\!\!{d_{k,n}} \!=\! {d_{k,n - 1}} - {v_{k,n - 1}}  \widetilde T\cos ( {\tilde \theta_{k,n-1}} ) \!+ \!u _{k,n }^d,\\
\!\!\!{v_{k,n}} \!=\! {v_{k,n - 1}} \!+ \!u _{k,n }^v,
%\\\alpha_{k,n}=\alpha_{k,{n-1}}\left( {1{\rm{ + 2}}\frac{{{v_{k,n - 1}}\cos \varphi \Delta T\cos \left( {{\theta _{k,n - 1}} - \varphi } \right)}}{{{d_{k,n - 1}}}}} \right)+\omega _{k,n }^\alpha
\end{array} \right.\label{eqARCI1}
\end{align}
where $u _{k,n }^\theta$, $u _{k,n }^d$, and $u _{k,n }^v$ are the corresponding evaluation noises on the angle, distance, and velocity, respectively. Here, $u _{k,n }^\theta$, $u _{k,n }^d$, and $u _{k,n }^v$ are assumed to follow Gaussian distributions with  means zero and variances $\delta_k^\theta$, $\delta_k^d$, and $\delta_k^v$, respectively.

From the approximated evaluation model in \eqref{eqARCI1},
the  radar channel aging model within one-block time aging can be rewritten as
\begin{align}
{{{\bf{x}}_{k,n}} = {\cal{G}}\left( {{{\bf{x}}_{k,n - 1}}} \right) + {{\bf{u }}_{k,n }} }\label{eqARCI2},
\end{align}
where ${\cal{G}}(\cdot)$ is the function dependent on \eqref{eqARCI1}, and
${{\bf{u}}_{k,n}}{\rm{ = }}{\left[ {u_{k,n}^\theta ,u_{k,n}^d,u_{k,n}^v} \right]^T} \in {{\mathbb C}^{3 \times 1}}$ is the noise with variance ${{\bm{\Sigma }}_{k}} = {\rm diag}\left(  {\delta _k^\theta ,\delta _k^d,\delta _k^v}  \right)\in{\mathbb C}^{3\times 3}$.

\subsubsection{  Communication Channel Aging Model}
Here, we assume that there is no frequency selectivity on the communication channels for simplicity \cite{gaudio2019performance}\footnote{
For the  frequency selective communication channels, we can modify the transmission frame structure during Phase-II to estimate the CSI on other subcarriers. Besides, as for exploring the effect of channel aging time on frequency-selective channels, we leave it as an interesting topic for future research.}. Let ${\bf h}_{q,n}\in {\mathbb C}^{L_t\times1}$ denote the downlink channel response from the BS to receiver ${U}_q^{\rm C}$ at the $n$-th block, which is assumed to follow Rayleigh fading, i.e., ${\bf h}_{q,0} \sim {\cal C}{\cal N}\left( {0,{\beta _q}{{\bf{I}}_{{L_t}}}} \right)$, where ${\beta _q}$ denotes the large-scale fading effect.
Then, the communication channel aging model within one-block time aging can be given by \cite{baddour2005autoregressive}, i.e.,
\begin{align}
{{\bf{h}}_{q,n}} = \rho_q {{\bf{h}}_{{q,n- 1} }} + \sqrt {1 - {\rho_q ^2}}  {\bf e}_{q,n},\label{eq10}
\end{align}
where ${\bm e}_{q,n}$ is the uncorrelated complex Gaussian noise at the $n$-th block with assuming ${\bm e}_{q,n} \sim {\cal CN}\left( {0,\beta_q {{\bf{I}}_{{L_t}}}} \right)$, and parameter ${\rho}_q$ is the temporal correlation coefficient dependent on the doppler shift ${\widetilde f_{q }}$ of receiver $U_q^{\rm C}$, which in Jakes' model is usually  given by $\rho_q  = {J_0}\left( {2\pi {\widetilde f_{q}}{\widetilde T}} \right)$ and ${J_0}(\cdot)$ is the zeroth-order Bessel function of the first kind.
Here, we assume that  the velocity of each user  keeps constant and the communication channel evolution is stationary. Thus, it can be off-line estimated by using the velocity estimator in \cite{schober2002velocity}.  Hence, $\rho_q$ can be assumed to be known as prior information at the BS \cite{deng2019intermittent}.

\section{ Communication-Radar Performance Analysis with Channel Aging}\label{secPerforemanceEvaluate}
In this section, we first analyze the radar tracking and channel estimation performances in the first block and then utilize the analyzed results to evaluate the tracking and communication performances within $n$-block time aging.

\subsection{Radar Performance with Channel Aging}
In this part, we first estimate the radar target mobility information and derive the corresponding CRLBs of estimated parameters in the first block. Then, we use the estimated results to evaluate the aged CRLBs in the $n$-th block without transmitting any training signals.

%predict the radar mobility information and further evaluate the corresponding CRLBs.

%\subsubsection{a}
%
% Then, the equivalent base band echo channel response between BS and target $U_k^{\rm R}$ is %given by
%\begin{align} {{\bf{Z}}_{k,n}}\left( {{{\bf{x}}_{k,n}},t,\tau } \right) = {\sqrt {{L_r}{L_t}} }{\alpha _{k,n}}e^{j\phi_k}{\bf{a}}\left( {{\theta _{_{k,n}}}} \right){{\bf{b}}^H}\left( {{\theta _{_{k,n}}}} \right)
%{e^{j2\pi \nu _{k,n}^Dt}}\delta \left( {\tau  - {\tau _{k,n}}} \right)\in {\mathbb C}^{L_r\times L_t}\label{eqradarchannel},
%\end{align}
%where $\delta(\cdot)$ is the Dirac delta function,

\subsubsection{Radar Channel Parameter Estimation}

In Phase-I of the first  block, by transmitting the training signals defined in \eqref{eqradartraining1},  the received reflected echoes at the BS on the $k$-th band through the round-trip radar channels are denoted by
${\bf{y}}_k^{\rm{R}}\left( t \right) \in {{\mathbb C}^{{L_r} \times 1}}$ \cite{gaudio2019performance}, i.e.,
%\begin{align}
%{\bf{y}}_k^{\rm{R}}\left( t \right) = \sum\limits_{k' = 1}^K \sqrt {{p_k}{L_r}{L_t}} {\alpha _{k',1}}{e^{j{\phi _{k'}}}}{\bf{a}}\left( {{\theta _{k',1}}} \right){{\bf{b}}^H}\left( {{\theta _{k',1}}} \right){{\bf{w}}_k}\left( t \right){e^{j2\pi \nu _{k',1}^Dt}}{s_k}\left( {t - {\tau _{k',1}}} \right) + {\bm \omega} _k^{\rm{R}}(t)  ,\label{eqisac1}
%\end{align}
\begin{align}
{\bf{y}}_k^{\rm{R}}\left( t \right) =& \sum\limits_{k' = 1}^K [\sqrt {{p_k}{L_r}{L_t}} {\alpha _{k',1}}{e^{j{\phi _{k'}}}}{\bf{a}}\left( {{\theta _{k',1}}} \right){{\bf{b}}^H}\left( {{\theta _{k',1}}} \right)\nonumber\\
&\times{{\bf{w}}_k}\left( t \right){e^{j2\pi \nu _{k',1}^Dt}}{s_k}\left( {t - {\tau _{k',1}}} \right)] + {\bm \omega} _k^{\rm{R}}(t)  ,\label{eqisac1}
\end{align}
where the upper subscript ``{$\rm R$}" implies the terms are related to the radar module and ${\bm{\omega }}_k^{\rm R} (t)$ is the received complex Gaussian noise at the BS with  power spectral density $ \sigma $. Here, ${\alpha _{k,1}} =  \sqrt {\frac{{{c_0^2}{\sigma _{{\rm{RCS,}}k}}}}{{{{\left( {4\pi } \right)}^3}f_c^2}{{d_{k,1}^{4}}}}} $,
${\tau _{k,1}} = \frac{{2{d_{k,1}}}}{{{c_0}}}$, and
$\nu _{k,1}^D = \frac{{2{v_{k,1}}}\cos \left( {{\theta _{k,1}} - \varphi_k } \right)}{c_0}{f_c}$  are  the attenuation factor due to  the propagation through the overall round-trip path  \cite{de2021joint,gaudio2020effectiveness}, time delay, and doppler phase shift, respectively, corresponding to target $U_k^{\rm R}$ at the $1$-th block.
Besides, ${\sigma _{{\rm{RCS,}}k}}$ and $\phi _k$ are the complex radar cross-section (RCS) coefficient and additional random phase noise of target $U_k^{\rm R}$, respectively.
Finally,  $c_0$ and $f_c$ are the speed of light and subcarrier frequency, respectively, and
${\bf{a}}\left( {{\theta }} \right)$ and ${{\bf{b}}}\left( {{\theta}} \right)$ are the angle-of-arrival (AoA) and  angle-of-departure (AoD) steering
vectors with respect to angle $\theta$, respectively. By setting the half-wavelength antenna spacing \cite{chen2019channel}, we have $
{\bf{a}}\left( \theta  \right){\rm{ = }}\frac{1}{{\sqrt {{L_r}} }}{\left[ {1,{e^{j\pi \sin \theta }}, \cdots ,{e^{j\pi \left( {{L_r} - 1} \right)\sin \theta }}} \right]^H}\in{\mathbb C}^{L_r\times 1}$ and $
{\bf{b}}\left( \theta  \right){\rm{ = }}\frac{1}{{\sqrt {{L_t}} }}{\left[ {1,{e^{j\pi \sin \theta }},\cdots,{e^{j\pi \left( {{L_t} - 1} \right)\sin \theta }}} \right]^H}\in{\mathbb C}^{L_t\times 1}$.

Let  ${{\bf{w}}_k}\left( t \right) = {\bf{b}}\left( {{{\hat \theta }_{k,1}^{\rm  P}}} \right)$ for $1\le k \le K$, where ${{{\hat \theta }_{k,1}^{\rm  P}}}$ is the predicted angle based on the initial radar target mobility state ${\bf x}_{k,0}$ using \eqref{eqARCI2}. Note that the initial state ${\bf x}_{k,0}$ is assumed to be known at the BS, which can be regarded as the aged mobility information in the last frame.
Then,  \eqref{eqisac1} can be equivalently rewritten   as
\begin{align}
\!\!\!{{\bf \hat y }_k^{\rm R}}\left( t \right) \!= \!{\alpha _{k,1}}{e^{j{\phi _k}}}{\bf{a}}\left( {{\theta _{k,1}}} \right){e^{j2\pi \nu _{k,1}^Dt}}{s_k}\left( {t - {\tau _{k,1}}} \right) +\hat {\bm  \omega}_k^{\rm R}(t) ,\label{eq11}
\end{align}
where  $\hat {\bm \omega}_k^{\rm R} (t)= \frac{\left( {{\bm{\omega }}_k^{\rm R}(t) + {{\bm \eta} _k}(t)} \right)}{{{\sqrt{p_k{L_r}{L_t}}}{\chi _{k,k}}}}$  is the interference-plus-noise term with ${\chi _{k,i}} = {{\bf{b}}^H}\left( {{\theta _{_{i,1}}}} \right){{\bf{b}}}\left( {{{\hat \theta }_{k,1}^{\rm  P}}}\right)$. Here, ${{\bm\eta} _k}(t){\rm{ = }}\sum\limits_{k' \ne k}^K {\sqrt{p_k}{\alpha _{k',1}}e^{j\phi_{k'}}{\bf{a}}\left( {{\theta _{k',1}}} \right){\chi _{k',k}}{e^{j2\pi \nu _{k',1}^Dt}}{s_k}\left( {t - {\tau _{k',1}}} \right)} $ is the interference caused by the reflected echoes of $s_k(t)$ from target ${U}_{k'}^{\rm R}$.
%, i.e., ${\bf u}_k^{\rm R} (t)\sim {\cal C}{\cal N}\left( {0,{ \bar\sigma}\Delta_f B_k{\bf I}_{L_r}} \right)$, where $\bar \sigma $  is the corresponding noise power spectral density.

 By sampling   ${{\bf \hat y }_k^{\rm R}}\left( t \right)$ in \eqref{eq11}  at each ${t = m{T }{\rm{ + }}{T_{cp}}{\rm{ + }}\frac{{{b}}}{{{B_k}}}T_o}$ after bandpass filtering and down conversion,   we have
\begin{align}
{{\tilde {\bf y}}_k^{\rm R}}\left( {m,{b}} \right)%& =
%{{\bf \hat y }_k^{\rm R}}\left( t \right) \left| {_{t = m{T }{\rm{ + }}{T_{cp}}{\rm{ + }}\frac{{{b}}}{{{B_k}}}T_o}} \right. \nonumber\\
&  \approx {\alpha _{k,1}}{e^{j{\phi _k}}}{\bf{a}}\left( {{\theta _{_{k,1}}}} \right){e^{j2\pi m{T}\nu _{k,1}^D}}\nonumber\\&
\!\!\!\!\!\!\! \times\sum\limits_{i = 0}^{{B_k} - 1} {\tilde s_{m,i}^k} {e^{j2\pi \frac{{i{b}}}{{{B_k}}}}}{e^{ - j2\pi i\Delta_f{\tau _{k,1}}}}
+{\hat {\bm \omega}_k^{\rm R}}(m,b),\label{eq12}
\end{align}
with assuming $\nu_{k,1}^{D} \ll  \Delta_f$, where ${\hat {\bm \omega}_{k}^{\rm R}}(m,b)$ is the sampled $\hat {\bm  \omega}_k^{\rm R}(t)$.
After applying the discrete Fourier transform (DFT), we have
\begin{align}
&{{\bf {\overline y}}_k^{\rm R}}\left( {m,{b}} \right) = \frac{1}{{\tilde s_{m,{b}}^k{B_k}}}{\sum\nolimits_{i' = 0}^{{B_k} - 1} {
{{\tilde {\bf y}}_k^{\rm R}}\left( {m,{i'}} \right)e} ^{ - j2\pi \frac{{i'{b}}}{{{B_k}}}}}\nonumber \\
&\approx { \alpha _{k,1}}{e^{j{\phi _k}}}{\bf{a}}\left( {{\theta _{_{k,1}}}} \right){e^{j2\pi m{T}\nu _{k,1}^D}}{e^{ - j2\pi {b}\Delta_f{\tau _{k,1}}}} + {\overline {\bm \omega}_{k}^{\rm R}}(m,b) ,\label{eqMeasurement1}
\end{align}
where ${\overline {\bm \omega }_{k}^{\rm R}}(m,b)$ is the corresponding noise.

From \eqref{eqMeasurement1}, the angle of $U_k^{\rm R}$ in the first block  can be estimated by
\begin{align}
 {{\hat \theta }_{_{k,1}}}{\rm{ = }}\mathop {\arg \max }\limits_\theta  \sum\nolimits_{m = 0}^{M_1 - 1} {\sum\nolimits_{{b} = 0}^{{B_k}-1} {{{\left| {{{\bf a}^H}\left( \theta  \right){{\bf{\overline y}}_k^{\rm R}}\left( {m,{b}} \right)} \right|}^2}} }.\label{eqa112}
\end{align}

Then, with the estimated angle ${{\hat \theta }_{_{k,1}}}$,  we apply the Periodogram-based method \cite{braun2014ofdm} to estimate the distance and velocity of $U_k^{\rm R}$.
Specifically, we define ${{\cal Y}_k}\left( \tau,{{\nu ^D} } \right)$ as
\begin{align}
&{{\cal Y}_k}\left( \tau,{{\nu ^D} }  \right) = \nonumber\\
&\sum\limits_{m = 0}^{M_1- 1} {\sum\limits_{{b} = 0}^{{B_k-1}} {{{\bf a}^H}\left( {{\hat \theta }_{_{k,1}}}  \right){{\bf{\overline y}}_k^{\rm R}}\left( {m,{b}} \right)}e^{ - j2\pi m{T}{\nu ^D}}{e^{j2\pi {b}\Delta_f\tau }}}.
\end{align}
Then, the time delay and doppler phase shift can be estimated by
\begin{align}\left( {{{ \hat \tau}_{k,1}}},{ \hat \nu} _{k,1}^D \right) = \mathop {\arg \max }\limits_{\tau,{{\nu ^D} }  } {\left| {{{\cal Y}_k}\left( \tau,{{\nu ^D} }  \right)} \right|^2}.
\end{align}
Finally, the distance and velocity of $U_k^{\rm R}$ in the first block can be estimated by
\begin{align}
{{\hat d}_{k,1}} &= \frac{1}{2}{{\hat \tau }_{k,1}}{c_0},\label{eqa115}\\
{{\hat v}_{k,1}}& {\rm{ = }}\frac{{\hat \nu _{k,1}^D{c_0}}}{{2{f_c}\cos \left( {{{\hat \theta }_{k,1}} - \varphi_k} \right)}}.\label{eqa116}
\end{align}
\subsubsection{Radar Tracking Performances in Block 1}

Here, we derive the CRLBs from the received signal model in \eqref{eqMeasurement1} for radar tracking performance analysis.
%\begin{theorem}\label{theorem000}
%Each   component  in  $\overline {\bm \omega} _k^{\rm{R}}(m,b)$ in \eqref{eqMeasurement1} approximately follows  Gaussian distribution with mean zero and variance  $\frac{{ \sigma  }\Delta_f}{p_kL_rL_t}$.
%\end{theorem}
%\begin{IEEEproof}
%a
%\end{IEEEproof}

To begin with, by applying some algebraic manipulations of ${{\chi _{k,i}}}$ in \eqref{eq11}, we have $\left| {{\chi _{k,i}}} \right| = \frac{{\left| {\sin \left( {\frac{{{L_t}\pi }}{2}\Delta _{k,i}^\theta } \right)} \right|}}{{{L_t}\left| {\sin \left( {\frac{\pi }{2}\Delta _{k,i}^\theta } \right)} \right|}}
$,
where $\Delta _{k,i}^\theta {{ = }} {\sin \left( {{\theta _{i,1}}} \right) - \sin \left( {\hat \theta _{k,1}^{\rm{P}}} \right)} $.
Then, we have $\left| {{\chi _{k,i}}} \right|=1$ if $\Delta _{k,i}^\theta=0$ and  $\left| {{\chi _{k,i}}} \right|=0$ if $-2<\Delta _{k,i}^\theta<2$ for large scale transmit antenna array, i.e., $L_t{ \to \infty }$.
Next, in order to  analyze the distribution of $\overline {\bm \omega} _k^{\rm{R}}(m,b)$ in \eqref{eqMeasurement1}, we apply the approximation that  $\left| {{\chi _{k,k}}} \right|=1$
 for  $1\le k    \le K$  and ${{\chi _{k,i}}}=0$   for $1\le k \ne i \le K$ by ignoring the bias of the predicted angle.
Note that  we only apply this approximation in performance analysis, and the non-zero bias exists in $\Delta _{k,k}^\theta$ and $\left| {{\chi _{k,k}}} \right|$ may be smaller than one in practice/simulations.  Nevertheless, this approximation only affects the noise power analysis but does no changes to the procedure of the proposed analysis method. From simulation results in Fig. \ref{figsimu1}, we observe that the non-zero bias does not influence the analyzed performance results much.
 Then, the noise ${\overline {\bm \omega}_{k}^{\rm R}}(m,b)$ in \eqref{eqMeasurement1} is approximated by%\footnote{Note that this assumption is only applied to simplify the remaining performance derivations, and $\Delta _{k,i}^\theta$ in the simulation is not equivalent to zero actually.}
 \begin{align}
\overline {\bm \omega} _k^{\rm{R}}(m,b) \approx  \frac{\sum\nolimits_{i' = 0}^{{B_k} - 1} {{\bm \omega}_k^{\rm{R}}\left( {m,b} \right){e^{ - j2\pi \frac{{i'b}}{{{B_k}}}}}}}{{\tilde s_{m,b}^k{B_k}}\sqrt{p_kL_rL_t}}  \label{eqnoiseterm},
\end{align}
where ${ {\bm \omega}_{k}^{\rm R}}(m,b)$ is the sampled $ {\bm  \omega}_k^{\rm R}(t)$.
Hence, we can know $\overline {\bm \omega} _k^{\rm{R}}(m,b)$ can be approximated as Gaussian distribution  with  mean zero and variance $ \frac{{ \sigma  }\Delta_f}{p_kL_rL_t} {\bf I}_{L_r}$.

Next, the CRLBs of radar target tracking in block 1 are given in the following theorem\footnote{
Note that applying other compressive algorithms may improve the estimation accuracy, but the derived results are the lower bounds of any unbiased estimators \cite{kay1993fundamentals}, which depend on the statics information in  \eqref{eqMeasurement1} instead of estimation algorithms. Hence, we do not provide other estimation algorithms due to the limited pages.}.
\begin{theorem}\label{theorem0}
In the regime of large $L_t$, the CRLBs on the estimation MSEs of ${\theta _{k,1}}$, distance ${d_{k,1}}$, and velocity ${v_{k,1}}$ for $U_k^{\rm R}$ in block 1  can be approximated by
\begin{subequations}\label{eqcrlb}
\begin{align}
{\mathbb E}\left( {{{\left\| {{\theta _{k,1}} - {{\hat \theta }_{k,1}}} \right\|}^2}} \right)& \ge\frac{{\sigma _k^\theta }}{{{p_k}{B_k}{M_1}}}%\buildrel \Delta \over ={\cal C}\left(\hat \theta_{k,1}\right)
\label{eqcrlb1},\\
{\mathbb E}\left( {{{\left\| {{d_{k,1}} - {{\hat d}_{k,1}}} \right\|}^2}} \right) &\ge\frac{{\sigma _k^d}}{{{p_k}{B_k}{M_1}\left( {B_k^2 - 1} \right)}}%\buildrel \Delta \over ={\cal C}\left(\hat d_{k,1}\right)
\label{eqcrlb2},\\
{\mathbb E}\left( {{{\left\| {{v_{k,1}} - {{\hat v}_{k,1}}} \right\|}^2}} \right) &\ge\frac{{\sigma _k^v}}{{{p_k}{B_k}{M_1} \left( {M_1^2 - 1} \right)}}%\buildrel \Delta \over ={\cal C}\left(\hat v_{k,1}\right)
\label{eqcrlb3},
\end{align}\end{subequations}
where $\sigma _k^\theta  = \frac{{6{ \sigma  }}{\Delta_f}}{{\left| {{\alpha _{k,1}}} \right|^2{\pi ^2}{{\cos }^2}\left( {{\theta _{k,1}}} \right)}L_tL_r\left( {L_{{r}}^2 - 1} \right)}$, $
\sigma _k^d = \frac{{3c_0^2{ \sigma  {\Delta_f}}}}{{8{{\left( {\pi \Delta_f} \right)}^2}\left| {{\alpha _{k,1}}} \right|^2}{L_tL_r}}$,
and $\sigma _k^v = \frac{{3c_0^2{ \sigma  {\Delta_f}} }}{{8{{\left( {\pi {T}} \right)}^2}f_c^2\left| {{\alpha _{k,1}}} \right|^2}{{\cos }^2}\left( \tilde \theta_{k,1} \right){L_tL_r}}$.
\end{theorem}
\begin{IEEEproof}
 Please refer to Appendix \ref{appCRLB}.
\end{IEEEproof}

\subsubsection{Aged Radar Tracking Performances in Block $n$}
In the following, we define the estimation/prediction error covariance matrix in the $n$-th block as
\begin{align}
{{\bf{E}}_{k,n}} ={\mathbb{E}}\left( {\left( {{{\bf{x}}_{k,n}} - {{{\bf{\hat x}}}_{k,n}}} \right){{\left( {{{\bf{x}}_{k,n}} - {{{\bf{\hat x}}}_{k,n}}} \right)}^H}} \right)\in{\mathbb C}^{3\times 3},\label{eqpredictionerror}
\end{align}
where ${{{\bf{\hat x}}}_{k,n}}$ is the aged radar mobility information in the $n$-th block.% for $2\le n \le N$.

From \eqref{eqARCI2}, the aged information ${{{\bf{\hat x}}}_{k,n}}$ can be predicted if ${{{\bf{\hat x}}}_{k,{n-1}}}$ is already known, i.e.,
\begin{align}
{{{\bf{\hat x}}}_{k,n}} = {\cal G}\left( {{{{\bf{\hat x}}}_{k,{n-1}}}} \right)\label{eqtrac1}.
\end{align}

However, since ${\cal G}(\cdot)$ defined in \eqref{eqARCI2} is a nonlinear function, it is difficult to derive the aged tracking performance analytically.
Thus, we apply the linear approximation for ${\cal G}(\cdot)$ in \eqref{eqARCI2}, i.e.,
\begin{align}
{{\bf{x}}_{k,n}}  \approx {\cal G}\left( {{{{\bf{\hat x}}}_{k,n - 1}}} \right) + {{\bf{G}}_{k,n - 1}}\left( {{{\bf{x}}_{k,n - 1}} - {{{\bf{\hat x}}}_{k,n - 1}}} \right) + {{\bf{u}}_{k,n}}\label{eqxlinear},
\end{align}
where $ {\bf{G}}_{k,{n-1}}\in{\mathbb C}^{3\times 3}$ is the Jacobian matrix for ${\partial {\cal G}\left( {{{\bf{x}}}_{k,{n-1}}} \right)}$, i.e.,  $ {{{\bf{G}}_{k,{n-1}}} = \frac{{\partial {\cal G}\left( {{{\bf{x}}_{k}}} \right)}}{{\partial {{\bf{x}}_{k}}}}\left| {_{{{\bf{x}}_{k}} = {{{\bf{\hat x}}}_{k,{n-1}}}}} \right.}$.

Next, by substituting \eqref{eqtrac1} and  \eqref{eqxlinear} into \eqref{eqpredictionerror}, matrix $
{{\bf{E}}_{k,n}}$ can be rewritten as
 \begin{align}
{{\bf{E}}_{k,n}} = & {\mathbb E}\left( {\left( {{{\bf{x}}_{k,n}} - {\cal G}\left( {{{{\bf{\hat x}}}_{k,n - 1}}} \right)} \right){{\left( {{{\bf{x}}_{k,n}} - {\cal G}\left( {{{{\bf{\hat x}}}_{k,n - 1}}} \right)} \right)}^H}} \right)\nonumber\\
%= &{\mathbb E}\left( {\left( {{{\bf{G}}_{k,n - 1}}\left( {{{\bf{x}}_{k,n - 1}} - {{{\bf{\hat x}}}_{k,n - 1}}} \right) + {{\bf{u}}_{k,n}}} \right){{\left( {{{\bf{G}}_{k,n - 1}}\left( {{{\bf{x}}_{k,n - 1}} - {{{\bf{\hat x}}}_{k,n - 1}}} \right) + {{\bf{u}}_{k,n}}} \right)}^H}} \right)\nonumber\\
 =  &{{\bf{G}}_{k,n - 1}}{{\bf{E}}_{k,n - 1}}{\bf{G}}_{k,n - 1}^H + {{\bf{\Sigma }}_{k} },
\end{align}
where ${{\bf{\Sigma }}_{k} }$ is defined in \eqref{eqARCI2}. After some  algebraic manipulations,
${{\bf{E}}_{k,n}}$  for $2\le n \le N$ can be rewritten as a function of ${{\bf{E}}_{k,1}}$, i.e.,
\begin{align}
{{\bf{E}}_{k,n}}
  = {{{\bf{\tilde G}}}_{k,n - 1}}{{\bf{E}}_{k,1}}{\bf{\tilde G}}_{k,n - 1}^H + \sum\limits_{i = 1}^{n - 1} {{{{\bf{\bar G}}}_{k,n,i}}} {{\bf{\Sigma }}_k }{{{\bf{\bar G}}}_{k,n,i}^H}\label{eq25a},
  \end{align}
where ${{{\bf{\tilde G}}}_{k,n - 1}}\in{\mathbb C}^{3\times 3}$ and ${{{\bf{\bar G}}}_{k,n,i}}\in{\mathbb C}^{3\times 3}$ are given by
\begin{align}
{{{\bf{\tilde G}}}_{k,n - 1}} = \prod\nolimits_{i = 1}^{n - 1} {{{\bf{G}}_{k,i}}} \;{\rm and}\;
{{{\bf{\bar G}}}_{k,n,i}} = \left\{ {\begin{array}{*{20}{c}}
{{\bf{I}}_3,i = 1,}\\
{\prod\nolimits_{i' = i}^{n - 1} {{{\bf{G}}_{k,i'}},i \ge 2,} }\label{eq26a}
\end{array}} \right.
\end{align}
respectively.
By using the CRLBs obtained in \eqref{eqcrlb}, we have
\begin{align}
&{{\bf{E}}_{k,1}} \nonumber\\&
\succeq {\rm{diag}}\left( {\frac{{\sigma _k^\theta }}{{{p_k}{B_k}{M_1}}},\frac{{\sigma _k^d}}{{{p_k}{B_k}{M_1}\left( {B_k^2 - 1} \right)}},\frac{{\sigma _k^v}}{{{p_k}{B_k}{M_1}\left( {M_1^2 - 1} \right)}}} \right)\nonumber\\&
\buildrel \Delta \over = {{\bf{D}}_k}.\label{eqcrlbo}
\end{align}
By substituting \eqref{eqcrlbo} into \eqref{eq25a}, we have
\begin{align}
{{\bf{E}}_{k,n}}
  &\succeq{{{\bf{\tilde G}}}_{k,n - 1}}{{\bf{D}}_{k}}{\bf{\tilde G}}_{k,n - 1}^H + \sum\nolimits_{i = 1}^{n - 1} {{{{\bf{\bar G}}}_{k,n,i}}} {{\bf{\Sigma }}_k }{{{\bf{\bar G}}}_{k,n,i}^H}\nonumber\\
&  \buildrel \Delta \over = {{\bf{\tilde E}}_{k,n}}\label{eq28b}.
  \end{align}
Thus, the aged CRLBs on the predicted MSEs of ${\theta _{k,n}}$, distance ${d_{k,n}}$, and velocity ${v_{k,n}}$ for $U_k^{\rm R}$ in block $ n$  are given by
\begin{subequations}\label{eq29t}
\begin{align}
{\mathbb E}\left( {\;{{\left\| {{\theta _{k,n}} - {{\hat \theta }_{k,n}}} \right\|}^2}} \right)& \ge {{\bf{\tilde E}}_{k,n}}(1,1)\buildrel \Delta \over ={\cal C}\left(\hat \theta_{k,n}\right),\label{eqcrlb1pa}\\
{\mathbb E}\left( {\;{{\left\| {{d_{k,n}} - {{\hat d}_{k,n}}} \right\|}^2}} \right) &\ge {{\bf{\tilde E}}_{k,n}}(2,2)\buildrel \Delta \over ={\cal C}\left(\hat d_{k,n}\right),\label{eqcrlb2pa}\\
{\mathbb E}\left( {\;{{\left\| {{v_{k,n}} - {{\hat v}_{k,n}}} \right\|}^2}} \right) &\ge {{\bf{\tilde E}}_{k,n}}(3,3)\buildrel \Delta \over ={\cal C}\left(\hat v_{k,n}\right).\label{eqcrlb3pa}
\end{align}
\end{subequations}

However, considering the matrix multiplication of polynomials in ${{{\bf{\tilde G}}}_{k,n - 1}}$ and ${{{\bf{\bar G}}}_{k,n,i}}$, it is still hard to apply the aged radar tracking performance in \eqref{eq29t} to study the resource allocation scheme design.  Therefore, we propose the following theorem to further approximate the CRLBs in \eqref{eq29t}, thus obtaining the simplified expressions.%, thus reducing the resource allocation design burden.
\begin{theorem}\label{theorem3}
The aged CRLBs on the predicted MSEs of ${\theta _{k,n}}$, distance ${d_{k,n}}$, and velocity ${v_{k,n}}$ for $U_k^{\rm R}$ in block $n$ derived in \eqref{eq29t} can be approximated by
\begin{subequations}\label{eq30t}
\begin{align}
{\cal C}\left( {{{\hat \theta }_{k,n}}} \right) &\approx \frac{{{a_{k,n}}\sigma _k^\theta }}{{{p_k}{B_k}{M_1}}} + {b_{k,n}}\delta _k^\theta \nonumber\\
& \buildrel \Delta \over = \widetilde {\cal C}\left( {{{\hat \theta }_{k,n}}} \right),\label{eqcrlb1p}\\
{\cal C}\left( {{{\hat d}_{k,n}}} \right) &\approx   \frac{{\sigma _k^d}}{{{p_k}{B_k}{M_1}\left( {B_k^2 - 1} \right)}} + \left( {n - 1} \right)\delta _k^d  \nonumber\\
& \buildrel \Delta \over = \widetilde {\cal C}\left( {{{\hat d}_{k,n}}} \right),\label{eqcrlb2p}\\
{\cal C}\left( {{{\hat v}_{k,n}}} \right) &\approx    {\frac{{\sigma _k^v}}{{{p_k}{B_k}{M_1}\left( {M_1^2 - 1} \right)}} + \left( {n - 1} \right)\delta _k^v} \nonumber\\
&   \buildrel \Delta \over = \widetilde {\cal C}\left( {{{\hat v}_{k,n}}} \right),\label{eqcrlb3p}
\end{align}
\end{subequations}
where ${a_{k,n}}={\left| {1 + \sum\nolimits_{i = 1}^{n - 1} {\frac{{{{\hat v}_{k,i}}  \widetilde T}}{{{{\hat d}_{k,i}}}}\cos ( {{\hat{\tilde \theta }_{k,i}}} )} } \right|^2}$, $b_{k,n} = 1 + \sum\nolimits_{i = 2}^{n - 1} {{{\left| {1 + \sum\nolimits_{i' = i}^{n - 1} {\frac{{{{\hat v}_{k,i'}}  \widetilde T}}{{{{\hat d}_{k,i'}}}}\cos ( {{\hat{\tilde \theta }_{k,i'}}} )} } \right|}^2}} $, and ${\hat{\tilde \theta }_{k,n}}={{\hat \theta }_{k,n}}{\rm{ - }}{\varphi _k}$.
\end{theorem}
\begin{IEEEproof}
Please refer to Appendix~\ref{appendixAppr1}.
\end{IEEEproof}
Note that from the simulation results in Fig. \ref{figsimu1}, we know that the proposed approximated CRLBs in \eqref{eq30t} can achieve nearly identical performances to that in \eqref{eq29t}.

\subsection{Communication Performance with Channel Aging}

In this part, we first estimate the CSI of all communication receivers in the first block and use them to predict the CSI in the $n$-th block. Then, we analyze the channel estimation/prediction error and derive the total aged achievable rate of all receivers.

\subsubsection{Communication CSI Estimation and Prediction}In Phase-I of the first transmission block, by transmitting the training signals in \eqref{eqradartraining1}, the received signals at $U_q^{\rm C}$ are given by
%\begin{align}
%y_{q,1}^{\rm{C}}\left( t \right) = &{\bf{h}}_{q,1}^H{{\bf{w}}_0}\left( t \right)\sum\nolimits_{m = 0}^{{M_1} - 1} {\sum\nolimits_{b = 0}^{{B_0} - 1} {\tilde s_{m,b}^0} } {e^{j2\pi b\Delta_f\left( {t - {T_{cp}} - m{T }} \right)}}%\nonumber\\&\times
%{\rm{rect}}\left( {t - m{T }} \right) + \omega_{q,1}^{\rm{C}}\left( t \right),\label{eq32}
%\end{align}
\begin{align}
y_{q,1}^{\rm{C}}\left( t \right) = &{\bf{h}}_{q,1}^H{{\bf{w}}_0}\left( t \right)\sum\limits_{m = 0}^{{M_1} - 1} {\sum\limits_{b = 0}^{{B_0} - 1} {\tilde s_{m,b}^0} } {e^{j2\pi b\Delta_f\left( {t - {T_{cp}} - m{T }} \right)}}\nonumber\\
&\qquad\qquad\qquad\times{\rm{rect}}\left( {t - m{T }} \right) + \omega_{q,1}^{\rm{C}}\left( t \right),\label{eq32}
\end{align}
where the upper subscript ``{$\rm C$}" implies the terms are related to the communication module. The term $\omega_{q,1}^{\rm{C}}\left( t \right)$ is the received Gaussian noise with power spectral density $\sigma _q^{\rm C}$.

Similar as the processes in \eqref{eq12} and \eqref{eqMeasurement1}, by sampling $y_{q,1}^{\rm{C}}\left( t \right)$ in \eqref{eq32}  at each ${t = m{T }{\rm{ + }}{T_{cp}}{\rm{ + }}\frac{{{b}}}{{{B_k}}}T_o}$ after bandpass filtering, and setting the corresponding transmit beamformer ${\bf w}_0(t)$ as ${\bf f}_0(m)$, we can apply the DFT for the sampled signals and have
% \begin{align}
%\!\!\!\!{{\bar y}}_{q,1}^{\rm{C}}\left( {m,b} \right)  = \frac{1}{{\tilde s_{m,b}^0{B_0}}}\sum\nolimits_{i' = 0}^{{B_0} - 1} {{{\widetilde y}}_q^{\rm{C}}\left( {m,i'} \right){e^{ - j2\pi \frac{{i'b}}{{{B_0}}}}}}
%  \approx \sqrt{p_0}{\bf{h}}_{q,1}^H{{\bf{f}}_0}\left( m \right) +  {{\bar \omega }}_{q,1}^{\rm{C}}\left( {m,b} \right),
% \end{align}
  \begin{align}
 {{\bar y}}_{q,1}^{\rm{C}}\left( {m,b} \right)  = &\frac{1}{{\tilde s_{m,b}^0{B_0}}}\sum\nolimits_{i' = 0}^{{B_0} - 1} {{{\widetilde y}}_q^{\rm{C}}\left( {m,i'} \right){e^{ - j2\pi \frac{{i'b}}{{{B_0}}}}}}\nonumber\\
  \approx &\sqrt{p_0}{\bf{h}}_{q,1}^H{{\bf{f}}_0}\left( m \right) +  {{\bar \omega }}_{q,1}^{\rm{C}}\left( {m,b} \right),
 \end{align}
where $  {{\bar \omega }}_{q,1}^{\rm{C}}\left( {m,b} \right)$ is the approximated Gaussian noise with  mean zero and variance ${{\tilde \sigma }_q}={ \sigma _q^{\rm C}}\Delta_f$.
Then,  during the $m$-th OFDM symbol duration, we have
 \begin{align}
{{\mathord{\buildrel{\lower3pt\hbox{$\scriptscriptstyle\smile$}}
\over y} }}_{q,1}^{\rm{C}}\left( m \right) &= \frac{1}{{{B_0}}}\sum\limits_{b = 1}^{{B_0}} {{{\bar y}}_{q,1}^{\rm{C}}\left( {m,b} \right)} \nonumber\\
 &\approx \sqrt{p_0}{\bf{h}}_{q,1}^H{{\bf{f}}_0}\left( m \right) + {{\mathord{\buildrel{\lower3pt\hbox{$\scriptscriptstyle\smile$}}\label{eq34f}
\over \omega } }}_{q,1}^{\rm{C}}\left( m \right),\end{align}
where  ${{\mathord{\buildrel{\lower3pt\hbox{$\scriptscriptstyle\smile$}}\label{eq34}
\over \omega } }}_{q,1}^{\rm{C}}\left( m \right)$ is the equivalent Gaussian noise with mean zero  and variance $\frac{{{\tilde \sigma }_q}}{B_0}$.

Then, denoting $
{\bf{\mathord{\buildrel{\lower3pt\hbox{$\scriptscriptstyle\smile$}}
\over y} }}_{q,1}^{\rm{C}} = {\left[ {\mathord{\buildrel{\lower3pt\hbox{$\scriptscriptstyle\smile$}}
\over y} _{q,1}^{\rm{C}}\left( 0 \right),\cdots,\mathord{\buildrel{\lower3pt\hbox{$\scriptscriptstyle\smile$}}
\over y} _{q,1}^{\rm{C}}\left( {{M_1} - 1} \right)} \right]^H}\in {\mathbb C}^{M_1\times 1}$ and $
{{\bf{F}}_0} = {\left[ {{{\bf f}_0}\left( 0 \right),\cdots,{{\bf f}_0}\left( {{M_1} - 1} \right)} \right]} \in {\mathbb C}^{L_t\times M_1}$,  we can rewrite \eqref{eq34f} as the following matrix form, i.e.,
 \begin{align}
{\bf{\mathord{\buildrel{\lower3pt\hbox{$\scriptscriptstyle\smile$}}
\over y} }}_{q,1}^{\rm{C}} = \sqrt{p_0}{{\bf{F}}_0^H}{\bf{h}}_{q,1} + {\bm{\mathord{\buildrel{\lower3pt\hbox{$\scriptscriptstyle\smile$}}
\over \omega } }}_{q,1}^{\rm{C}},
\end{align}
where  ${\bm{\mathord{\buildrel{\lower3pt\hbox{$\scriptscriptstyle\smile$}}
\over \omega } }}_{q,1}^{\rm{C}}\in{\mathbb C}^{M_1\times 1}$ is the received  noise, i.e., ${\bm{\mathord{\buildrel{\lower3pt\hbox{$\scriptscriptstyle\smile$}}
\over \omega } }}_{q,1}^{\rm{C}} \sim {\cal C}{\cal N}\left( {0,\frac{{{\tilde \sigma }_q}}{B_0}{{\bf I}_{{M_1}}}} \right)$.

By setting ${{\bf{F}}_0}$ satisfying ${{\bf{F}}_0}{\bf{F}}_0^H = \frac{M_1}{{{L_t}}}{{\bf{I}}_{{L_t}}}$, the linear MMSE channel estimator is given by \cite{papazafeiropoulos2016impact}
\begin{align}{{{\bf{\hat h}}}_{q,1}} = {\left( {1 +{\frac{{{L_t}{{\tilde \sigma }_q}}}{{{M_1}{B_0}{p_0}}}}} \times\frac{1}{{\beta _q}}\right)^{ - 1}}\left(\frac{1}{{\sqrt {{p_0}} }}{\bf F}_0^\dag {\bf{\mathord{\buildrel{\lower3pt\hbox{$\scriptscriptstyle\smile$}}
\over y} }}_{q,1}^{\rm{C}}\right)\label{eq36},
\end{align}
where ${\bf F}_0^\dag$ is the Moore-Penrose pseudoinverse of matrix ${\bf F}_0$.

From \eqref{eq36}, it can be proved that ${{{\bf{\hat h}}}_{q,1}} \sim {\cal C}{\cal N}\left( 0,{\lambda_q}{\bf I}_{L_t} \right)$ where ${\lambda _q} = \frac{{{M_1}\beta _q^2{p_0}{B_0}}}{{{L_t}{{\tilde \sigma }_q}  + {M_1}{\beta _q}{p_0}{B_0}}}$.
By using the orthogonality property of MMSE, channel ${\bf h}_{q,1}$ can be represented by
\begin{align}
{\bf h}_{q,1}={{{\bf{\hat h}}}_{q,1}}+{{{\bf{\tilde h}}}_{q,1}}\label{eq37a},
\end{align}
where ${{{\bf{\tilde h}}}_{q,1}} \sim {\cal C}{\cal N}\left( {0, \left( {\beta_q - {\lambda _q}} \right)}{\bf I}_{L_t} \right)$ is the corresponding channel estimation error.

By applying algebraic manipulations with the channel aging model in \eqref{eq10},  ${\bf h}_{q,n}$ is rewritten by
 \begin{align}
{{\bf{h}}_{q,n}} &= {\rho_q ^{n - 1}}{{\bf{h}}_{q,1}} + \sqrt {1 - {\rho_q ^{2(n - 1)}}} {{{\bf{\tilde e}}}_{q,n}},%\nonumber\\
\label{eqchannelpre}
\end{align}
where ${{{\bf{\tilde e}}}_{q,n}} = \frac{1}{{\sqrt {1 - \rho _q^{2(n - 1)}} }}\sum\limits_{i = 0}^{n - 2} {\rho _q^i\sqrt {1 - \rho _q^2} {{\bf{e}}_{q,n - i}}} $ is called  aging noise, which follows complex Gaussian
 distribution, i.e., ${{\bf{\tilde e}}}_{q,n}\sim {\cal C}{\cal N}\left( {0,\beta_q{{\bf{I}}_{{L_t}}}} \right) $.

Then, by substituting \eqref{eq37a} into \eqref{eqchannelpre},  we have
 \begin{align}
{{\bf{h}}_{q,n}}  &= {\rho_q ^{n - 1}}{{{\bf{\hat h}}}_{q,1}} + {\rho_q ^{n - 1}}{{{\bf{\tilde h}}}_{q,1}} + \sqrt {1 - {\rho_q ^{2(n - 1)}}} {{{\bf{\tilde e}}}_{q,n}}\label{eqchannelpre02}.
\end{align}
The predicted CSI of receiver $U_q^{\rm R}$ in the $n$-th block can be given by \cite{deng2019intermittent}
 \begin{align}
{{{\bf{\hat h}}}_{q,n}} = {\rho_q ^{n-1}}{{{\bf{\hat h}}}_{q,1}},\label{eq40}
\end{align}
whose prediction error is ${{{\bf{\bar e}}}_{q,n}} = {\rho_q ^{n - 1}}{{{\bf{\tilde h}}}_{q,1}} + \sqrt {1 - {\rho_q ^{2(n - 1)}}} {{{\bf{\tilde e}}}_{q,n}} $, which follows complex Gaussian distribution, i.e., ${{{\bf{\bar e}}}_{q,n}}\sim  {\cal C}{\cal N}\left( {0,\left({{\beta _q} - {\rho_q ^{2n - 2}}{\lambda _q} }\right){{\bf{I}}_{{L_t}}}} \right)$.

\subsubsection{Total Aged Achievable Rate}
Based on the data signals defined in \eqref{eq3}, the received signals at $U_q^{\rm C}$ in the $n$-th block can be rewritten as
\begin{align}
 &{y_{q,n}^{\rm C}}\left( t \right)% \nonumber\\=&
 = {\bf{h}}_{q,n}^H\sum\limits_{i = 1}^Q {\sqrt {{{\tilde p}_{i,n}}}{{\bf{f}}_{i,n}}{c_{i,n}}\left( t \right)}  + { \omega_{q,n}^{\rm C}}\left( t \right)\nonumber\\
& =   \sqrt {{{\tilde p}_{q,n}}} { {\mathbb E}}\left( {{\bf{h}}_{q,n}^H{{\bf{f}}_{q,n}}} \right){c_{q,n}}\left( t \right) + {\zeta _{q,n}}\left( t \right) + {{\tilde \zeta }_{q,n}}\left( t \right) + { \omega _{q,n}^{\rm C}}\left( t \right),
\end{align}
where  ${\omega_{q,n}^{\rm C}}\left( t \right)$ is the received Gaussian noise with  mean  zero and variance ${{\tilde \sigma }_q}$, and interference terms $ {\zeta _{q,n}}\left( t \right) $ and $ {{\tilde \zeta }_{q,n}}\left( t \right) $ are given by
 \begin{align}
{\zeta _{q,n}}\left( t \right)& = \sqrt {{{\tilde p}_{q,n}}} \left[ {{\bf{h}}_{q,n}^H{{\bf{f}}_{q,n}} - {\mathbb E}\left( {{\bf{h}}_{q,n}^H{{\bf{f}}_{q,n}}} \right)} \right]{c_{q,n}}\left( t \right),\\
{{\tilde \zeta }_{q,n}}\left( t \right) & = {\bf{h}}_{q,n}^H\sum\limits_{i \ne q}^Q {\sqrt {{{\tilde p}_{i,n}}} {{\bf{f}}_{i,n}}{c_{i,n}}\left( t \right)}\label{eqsinr0},
\end{align}
respectively.
Then, the achievable rate (bits/s/Hz) of  $U_q^{\rm C}$ in the $n$-th block is
\begin{align}
{R_{q,n}} = \frac{{{M_{n}'}}}{M}{\log _2}\left( {1 + {{\tilde p}_{q,n}}{\gamma _{q,n}}} \right),
\end{align}
where $M_1'=M-M_1$,  $M_n'=M$ for $2\le n\le N$, and
\begin{align} {\gamma _{q,n}} = \frac{{{{\left| {{\mathbb{E}}\left( {{\bf{h}}_{q,n}^H{{\bf{f}}_{q,n}}} \right)} \right|}^2}}}{{{\mathbb{E}}\left( {{{\left| {{\zeta _{q,n}}\left( t \right)} \right|}^2}} \right) + {\mathbb{E}}\left( {{{\left| {{{\tilde \zeta }_{q,n}}\left( t \right)} \right|}^2}} \right) + {\tilde \sigma _q}}}\label{eqsinr}.
\end{align}

Based on the predicted CSI in \eqref{eq40},  the MRT and ZF transmit beamforming vectors for each $U_q^C$ are given by
\begin{align}
{{\bf{f}}_{q,n}} = \left\{ {\begin{array}{*{20}{l}}
{\frac{{{{{\bf{\hat h}}}_{q,1}}}}{{\sqrt {{\mathbb E}\left( {{{\left\| {{{{\bf{\hat h}}}_{q,1}}} \right\|}^2}} \right)} }} = \frac{1}{{\sqrt {{\lambda _q}{L_t}} }}{{{\bf{\hat h}}}_{q,1}},{\rm{MRT}}},\\
{\frac{{{{\bf{a}}_q}}}{{\sqrt {{\mathbb E}\left( {{{\left\| {{{\bf{a}}_q}} \right\|}^2}} \right)} }}\mathop  = \limits^{(a)} \sqrt {{\lambda _q}\left( {{L_t} - Q} \right)} {{\bf{a}}_q},{\rm{ZF}}},
\end{array}} \right.\label{eqzfmrtBeam}
\end{align}
respectively, where ${\bf a}_q$ is the $q$-th column of matrix ${{{\bf{\hat H}}}_1}{\left( {{\bf{\hat H}}_1^H{{{\bf{\hat H}}}_1}} \right)^{ - 1}}$ with ${{{\bf{\hat H}}}_1} = \left[ {{{{\bf{\hat h}}}_{1,1}},{{{\bf{\hat h}}}_{2,1}},\cdots,{{{\bf{\hat h}}}_{Q,1}}} \right]\in\mathbb C ^{L_t\times Q}$ and (a) is due to
${\mathbb  E}({\left\| {{{\bf{a}}_q}} \right\|^2}) = \frac{1}{{{\lambda _q}\left( {{L_t} - Q} \right)}}$ for the scenario  $L_t\ge( Q+1)$ \cite{ngo2013energy}.
\begin{theorem}\label{Theorem31}
In the $n$-th aged block, with the predicted CSI in \eqref{eq40}, parameter $\gamma_{q,n}$ in \eqref{eqsinr} using MRT or ZF transmit beamforming  is given by
\begin{align}
 {\gamma _{q,n}}  
=  \left\{ {\begin{array}{*{20}{l}}
{\frac{{{\rho_q ^{2\left( {n - 1} \right)}}{M_1}\beta _q^2{p_0}{B_0}{L_t}}}{{\left( {P{\beta _q} + {\tilde \sigma _q}} \right)\left( {{L_t}{{\tilde \sigma }_q} + {M_1}{\beta _q}{p_0}{B_0}} \right)}},{\rm{MRT}}},\\
{\frac{{{\rho_q ^{2\left( {n - 1} \right)}}{M_1}\beta _q^2{p_0}{B_0}\left( {{L_t} - Q} \right)}}{{\left( {P{\beta _q} + {\tilde \sigma _q}} \right)\left( {{L_t}{{\tilde \sigma }_q} + {M_1}{\beta _q}{p_0}{B_0}} \right) - {\rho_q ^{2\left( {n - 1} \right)}}P{M_1}\beta _q^2{p_0}{B_0}}},{\rm{ZF}}}.
\end{array}} \right.\label{eqrateZFMRT}
\end{align}
\end{theorem}
\begin{IEEEproof}
 Please refer to Appendix \ref{appendixTheorem31}.
\end{IEEEproof}

Let the total achievable rate in the $n$-th block be ${R_n} = \sum\nolimits_{q = 1}^Q {{R_{q,n}}} $. Based on Theorem \ref{Theorem31},
 it is straightforward to have the following asymptotic analysis of   ${R_n} $  with respect to the transmit power, i.e.,
\begin{align}
&\mathop {\lim }\limits_{P \to \infty } {R_n} \to \nonumber\\
&\left\{ {\begin{array}{*{20}{l}}
{\frac{{{M_{n}'}}}{M}\sum\limits_{q = 1}^Q {{{\log }_2}\left( {1 + \rho _q^{2\left( {n - 1} \right)}{\kappa _{q,n}}{L_t}} \right)} ,{\rm{MRT}},}\\
{\frac{{{M_{n}'}}}{M}\sum\limits_{q = 1}^Q {{{\log }_2}\left( {1 + \frac{{\rho _q^{2\left( {n - 1} \right)}}}{{1 - \rho _q^{2\left( {n - 1} \right)}}}{\kappa _{q,n}}\left( {{L_t} - Q} \right)} \right)} ,{\rm{ZF}},}
\end{array}} \right.\label{eq47a}
\end{align}
where  ${\kappa _{q,n}} = \frac{{{{\tilde p}_{q,n}}}}{P}$. From \eqref{eq47a}, it can be observed that the achievable rates under MRT and ZF beamforming decrease as the channel aging time increases.

\section{Average Total Achievable Rate Maximization with Channel Aging}
In this section, we use the characterized aged radar tracking and communication performance metrics to study the situation-dependent resource allocation scheme considering the individual target tracking precision demand and communication rate requirement. In particular, we first provide the problem formulation and then provide the suboptimal algorithm to solve it efficiently.

\subsection{Problem Formulation}

From \eqref{eq30t} and \eqref{eqrateZFMRT}, besides the channel aging time, we know bandwidth and power allocations are also key factors that affect the aged communication and radar tracking performance.
Therefore, in this paper, we aim to jointly optimize the number of subcarriers ${\bf{B}} = \left[ {{B_0},\cdots,{B_K}} \right]^{T} \in {{\mathbb C}^{\left( {K + 1} \right)\times 1 }}$ and transmit power ${\bf{p}} = \left[ {{p_0},\cdots,{p_K}} \right]^{T} \in {{\mathbb C}^{\left( {K + 1} \right) \times 1}}$ for channel training in Phase-I of the first block, transmit power ${\bf{\tilde p}} = \left[ {{\tilde p_{1,1}},\cdots,{\tilde p_{q,n}},\cdots,{\tilde p_{Q,N}}} \right]^{T} \in {{\mathbb C}^{QN \times 1}}$ for data transmission in Phase-II of the first block and remaining blocks, and channel re-estimation interval $N$, to maximize the average total achievable rate subject to various practical constraints. Mathematically, we have the following optimization problem:
\begin{subequations}\label{eqProblemFromulation}
\begin{align}
{\bf P1}:\quad \mathop {\max }\limits_{N,{\bf{p}},\widetilde{\bf{ p}},{\bf{B}}}\;\;&{\cal R}\left( {N,{\bf{p}},{\bf{\tilde p}},{\bf{B}}} \right)= \frac{1}{N}\sum\nolimits_{n = 1}^N {\sum\nolimits_{q = 1}^Q {{R_{q,n}}} }  \nonumber\\
\quad{\rm {s.t.}}\quad&{\rm C1:}\; {\widetilde {\cal C}}\left( {{{\hat \theta }_{k,n}}} \right) \le \Gamma _{k,{\rm max}}^\theta, \; \forall  n, 1\le k\le K, \nonumber\\%\label{eqProblemFromulationa}\\
&{\rm C2:}\; {\widetilde {\cal C}}\left( {{{\hat d}_{k,n}}} \right) \le \Gamma _{k,{\rm max}}^d, \; \forall n,  1\le k\le K,\nonumber\\%\label{eqProblemFromulationb}\\
&{\rm C3:}\; {\widetilde{ \cal C}}\left( {{{\hat v}_{k,n}}} \right) \le \Gamma _{k,{\rm max}}^v, \; \forall n,  1\le k\le K,\nonumber\\%\label{eqProblemFromulationc}\\
&{\rm C4:} \;{R_{q,n}} \ge {R_{q,{\rm min}}}, \forall q, n,\nonumber\\%\label{eqProblemFromulationcc}\\
&{\rm C5:}\; \sum\nolimits_{k = 0}^K {{p_k}}  \le P, \; \nonumber\\%\label{eqProblemFromulatione}\\
&{\rm C6:} \;\sum\nolimits_{q = 1}^Q {{{\tilde p}_{q,n}}}  \le P, \; \forall n,\nonumber\\%\label{eqProblemFromulationd} \\
&{\rm C7:} \;\sum\nolimits_{k = 0}^K {{B_k}}  \le B, \;\nonumber\\%\label{eqProblemFromulationf} \\
&{\rm C8:}\; 0 \le {p_k}, \forall k,\nonumber\\%
&{\rm C9:}\; 0 \le {{{\tilde p}_{q,n}}}, \forall q,n,\nonumber\\%\label{eqProblemFromulationg}\\
&{\rm C10:} \;{B_k}  \in {\mathbb N}, \; \forall k,\nonumber%\label{eqProblemFromulationh}
\end{align}
\end{subequations}
where  $\Gamma_{k,{\rm max}}^\theta$, $\Gamma_{k,{\rm max}}^d$, and $\Gamma_{k,{\rm max}}^v$ are tracking error constraints of target $U_k^{\rm R}$ with respect to its angle, distance, and velocity, respectively, while
${R_{q,{\rm min}}}$ is the minimum transmission rate required by receiver $U_q^{\rm C}$. Specifically,
$\rm C1$, $\rm C2$, and $\rm C3$ are the individual radar tracking accuracy requirements and $\rm C4$ is the individual communication data rate requirement. $\rm C5$ and $\rm C6$ are the total transmission power constraints in the training and data transmission stages, respectively. $\rm C7$ is the total number of subcarriers constraint. Finally, $\rm C8$ , $\rm C9$, and $\rm C10$  are practical constraints.
%
%${\bf{p}} = \left[ {{p_0},\cdots,{p_K}} \right]^{T} \in {{\mathbb C}^{\left( {K + 1} \right) \times 1}}$, ${\bf{\tilde p}} = \left[ {{p_{1,1}},\cdots,{p_{q,n}},\cdots,{p_{Q,N}}} \right]^{T} \in {{\mathbb C}^{QN \times 1}}$, and  ${\bf{B}} = \left[ {{B_0},.,{B_K}} \right]^{T} \in {{\mathbb C}^{\left( {K + 1} \right) \times 1}}$.

\subsection{Algorithm Development}
%In this part, we propose an efficient suboptimal algorithm to solve problem ${\bf P1}$.

To begin with, considering integer constraint $\rm C10$, problem ${\bf P1}$ is MINLP in general. Thus, it is a non-convex optimization problem, and obtaining its optimal solution may introduce intractable computational complexity. In the following,  we  propose an efficient two-step based suboptimal algorithm to solve problem ${\bf P1}$.
Specifically, in the first step, we solve the relaxed continuous variable optimization problem of  ${\bf P1}$ by replacing constraint $\rm C10$ as $0\le{B_k},  \forall k$, i.e.,
\begin{align}
{\bf P2}:\quad\mathop {\max }\limits_{N,{\bf p},{\bf{\tilde p}},{\bf B} }\qquad&
{\cal R}\left( {N,{\bf{p}},{\bf{\tilde p}},{\bf{B}}} \right) \nonumber\\
{\rm {s.t.}}\qquad& 0\le{B_k} ,  \forall k, \label{eqp21}\\
& {\rm C1}-{\rm C8},\nonumber
\end{align}
Then, in the second step, we apply an integer conversion method to make the solutions of ${B_k}$  satisfy C10. Note that the optimal solution of problem ${\bf P2}$  is the upper bound of the original problem ${\bf P1}$ because the relaxed constraints can introduce a large feasibility region.

However, ${\bf P2}$ is still non-convex due to the coupled optimization variables. Furthermore, the numbers of optimization variables and conditions vary with the channel aging time $N$, which increases the difficulties in solving this problem. Therefore, to find the optimal solution of ${\bf P2}$,  we propose a one-dimensional search based optimization algorithm thanks to the channel aging time $N$ belonging to one-dimensional variable. In this algorithm, we use a one-dimensional exhaustive search for all possible channel aging time $N$, and we propose algorithms to solve ${\bf P2}$ optimally for each fixed $N$. Then, we can obtain the optimal solution of ${\bf P2}$ with the optimal $N$ that makes the objective maximum.

Next, in the following two theorems, we first analyze the potential upper bound of channel aging time $N$ to reduce search region in the one-dimensional search for complexity reduction, and then provide the way to find the optimal solution of ${\bf P2}$ with a fixed $N$.
\begin{theorem}\label{Theorem4a}
If problem ${\bf P2}$ is feasible,  the maximum potential feasible channel aging time $N$ should be smaller than the following bound, i.e.,
 \begin{align}
{N_{\max }} = \mathop {\min }\limits_{\forall k,\forall q} \left\{ {N_{k,\max }^{\rm{R}},N_{q,\max }^{\rm{C}}} \right\},\label{eqNmaxa}
\end{align}
where $N_{k,\max }^{\rm R}$ and $N_{k,\max }^{\rm C}$ are the potential maximum channel aging time $N$  for target $U_k^{\rm R}$  and receiver $U_k^{\rm C}$, respectively, which are given by
%\begin{small}
\begin{align}
\!\!N_{k,\max }^{\rm{R}} &\!=\!1+\!  \min \left\{ {\left\lfloor {\frac{{\Gamma _{k,{\rm{max}}}^d}}{{\delta _k^d}} - \frac{{\sigma _k^d}}{{\delta _k^d}{PB{M_1}\left( {{B^2} - 1} \right)}}} \right\rfloor ,} \right.\nonumber\\
&\qquad\qquad\left. {\left\lfloor {\frac{{\Gamma _{k,{\rm{max}}}^v}}{{\delta _k^v}} - \frac{{\sigma _{k}^v}}{{\delta _k^v}{PB{M_1}\left( {M_1^2 - 1} \right)}}} \right\rfloor } \right\}, \label{eqtheo1}\\
N_{q,\max }^{\rm{C}}{\rm{ }}& = \left\{ {\begin{array}{*{20}{l}}
{\left\lfloor {\frac{1}{{2\ln {\rho _q}}}\ln \left( {\frac{{\psi _{q,2}^{{\rm{MRT}}}}}{{\psi _{q,1}^{{\rm{MRT}}}}}\left( {{2^{{R_{q,\min }}}} - 1} \right)} \right)} \right\rfloor  + 1,{\rm{MRT}},}\\
{\left\lfloor {\frac{1}{{2\ln {\rho _q}}}\ln \left( {\frac{{\left( {{2^{{R_{q,\min }}}} - 1} \right)\psi _{q,2}^{{\rm{ZF}}}}}{{\psi _{q,1}^{{\rm{ZF}}}{\rm{ + }}\left( {{2^{{R_{q,\min }}}} - 1} \right)\psi _{q,3}^{{\rm{ZF}}}}}} \right)} \right\rfloor  + 1,{\rm{ZF}},}\label{eqtheo2}
\end{array}} \right.
\end{align}
%\end{small}
$\!\!\!$respectively, where  $\psi _{q,1}^{{\rm{MRT}}} = {M_1}\beta _q^2P^2B{L_t}$, $
\psi _{q,2}^{{\rm{MRT}}} = \left( {P{\beta _q} + {{\tilde \sigma }_q}} \right)\left( {{L_t}{{\tilde \sigma }_q} + {M_1}{\beta _q}PB} \right)$
, $\psi _{q,1}^{{\rm{ZF}}} = {M_1}\beta _q^2P^2B\left( {{L_t} - Q} \right)$, $
\psi _{q,2}^{{\rm{ZF}}} = \left( {P{\beta _q} + {{\tilde \sigma }_q}} \right)\left( {{L_t}{{\tilde \sigma }_q} + {M_1}{\beta _q}PB} \right)$
, and $\psi _{q,3}^{{\rm{ZF}}} = {M_1}\beta _q^2P^2B$.
\end{theorem}
\begin{IEEEproof}
Firstly, it is obviously that when all the power and bandwidth are allocated to $U_k^{\rm R}$, it can achieve the largest channel aging time under the target tracking conditions, i.e., $p_k=P$ and $B_k=B$. Then, substituting \eqref{eq30t} into constraints ${\rm C2}$ and ${\rm C3}$, we have
 \eqref{eqtheo1}. Similarly, by substituting \eqref{eqrateZFMRT} into constraints ${\rm C4}$ with assuming $p_0B_0=PB$, we have \eqref{eqtheo2}.
% \begin{align}
%\delta _k^d\left( {n - 1} \right) \le \Gamma _{k,{\rm{max}}}^d - \frac{{\sigma _k^d}}{{PB{M_1}\left( {{B^2} - 1} \right)}},\\
%\delta _k^v\left( {n - 1} \right) \le \Gamma _{k,{\rm{max}}}^v - \frac{{\sigma _k^v}}{{PB{M_1}\left( {M_1^2 - 1} \right)}}.
%\end{align}
\end{IEEEproof}
\begin{theorem} \label{theoremtransform}
If problem $\bf {P2}$ with the fixed $N$ is feasible, its optimal solution can  be obtained by successively solving the following two subproblems, i.e.,
\begin{align}
{\bf P2-A}:\;\left( {{{\bf{p}}^\star},{{\bf{B}}^\star}} \right)= &\mathop {\arg \max }\limits_{{\bf{p}},{\bf{B}}}\;p_0B_0  \nonumber\\
 &{\rm {s.t.}}\; {\rm C1}-{\rm C3},\;{\rm C5},\;{\rm C7},\;{\rm C8},\; {\rm and} \;\eqref{eqp21}, \nonumber
\end{align}
 and
\begin{align}
{\bf P2-B}:\;  {{{\bf{\tilde p}}^\star}} = &\mathop {\arg \max }\limits_{{\bf{\tilde p}}} \;   {\cal R}\left( {N,{{\bf{p}}^{\star}},{\bf{\tilde p}},{{\bf{B}}^{\star}}} \right)\nonumber\\
 &{\rm {s.t.}}\; {\rm C4},\;{\rm C6},\; {\rm and} \;{\rm C9},\nonumber
\end{align}
where  ${{\bf{p}}^\star}$ and ${{\bf{B}}^\star}$ are obtained by solving problem ${\bf P2-A}$.
\end{theorem}
\begin{IEEEproof}
Please refer to Appendix \ref{theoremtransformapp}.
\end{IEEEproof}

\subsection{Optimal Solution to Problem ${\bf {P2-A}}$}
In this part, we provide the optimal solution to problem ${\bf {P2-A}}$.

With the derived CRLBs in \eqref{eq30t}, constraints ${\rm C1}$, ${\rm C2}$, and ${\rm C3}$ can be equivalently rewritten as
\begin{align}
&{p_k}{B_k} \ge \tilde \Gamma _{k,N}^{\theta ,v},1\le k\le K\label{eq52},\\
&{p_k}{B_k}\left( {B_k^2 - 1} \right) \ge \tilde \Gamma _{k,N}^d,1\le k\le K \label{eq53},
\end{align}
where
%\begin{align}
%\tilde \Gamma _{k,N}^{\theta ,v} & = \max \left\{ {\left\{ {\frac{{{a_{k,n}}\sigma _k^\theta }}{{\left( {\Gamma _{k,\rm max}^\theta  - {b_{k,n}}\delta _k^\theta } \right){M_1}}},\forall n} \right\},\frac{{\sigma _k^v}}{{{M_1}({M_1^2-1})\left( {\Gamma _{k,\rm max}^v  - \left( {N - 1} \right)\delta _k^v} \right)}}} \right\},\\
%\tilde
%\Gamma _{k,N}^d & = \frac{{\sigma _k^d}}{{{M_1}\left( {\Gamma _{k,\rm max}^d  - \left( {N - 1} \right)\delta _k^d} \right)}}.
%\end{align}
\begin{align}\tilde \Gamma _{k,N}^{\theta ,v}& = \max \left\{ \begin{array}{l}
\left\{ {\frac{{{a_{k,n}}\sigma _k^\theta }}{{\left( {\Gamma _{k,{\rm{max}}}^\theta  - {b_{k,n}}\delta _k^\theta } \right){M_1}}},\forall n} \right\},\\
\frac{{\sigma _k^v}}{{{M_1}(M_1^2 - 1)\left( {\Gamma _{k,{\rm{max}}}^v - \left( {N - 1} \right)\delta _k^v} \right)}}
\end{array} \right\},\\
 \tilde\Gamma _{k,N}^d & = \frac{{\sigma _k^d}}{{{M_1}\left( {\Gamma _{k,\rm max}^d  - \left( {N - 1} \right)\delta _k^d} \right)}}.
\end{align}

Then, problem $\bf {P2-A}$ can be equivalently rewritten as
\begin{subequations}\label{eqP2A1}
\begin{align}
&\!\!\!\!\!\!{\bf {P2-A1}}:\mathop {\max }\limits_{{\bf{p}},{\bf{B}}} \;\;\ln \left( {{p_0}} \right) + \ln \left( {{B_0}} \right) \nonumber\\
{\rm {s.t.}}&\;\;\ln \left( {{p_k}} \right) + \ln \left( {{B_k}} \right) \ge \ln\left(\tilde \Gamma _{k,N}^{\theta ,v}\right),1 \le k \le K,\label{eqP2A1a}\\
&\;\;\ln \left( {{p_k}} \right) + \ln \left( {{B_k}} \right) + \ln \left( {{B_k} + 1} \right) \nonumber\\
&\;\;  + \ln \left( {{B_k} - 1} \right) \ge \ln\left(\tilde \Gamma _{k,N}^d\right),1 \le k \le K,\label{eqP2A1b}\\
&\;\;{\rm C5},\;{\rm C7},\;{\rm C8},\; {\rm and} \;\eqref{eqp21}, \nonumber
\end{align}
\end{subequations}
where constraints \eqref{eqP2A1a} and \eqref{eqP2A1b} are due to \eqref{eq52} and \eqref{eq53}.
It is straightforward to know that problem $\bf {P2-A1}$ is a convex optimization problem, and its optimal solution can be efficiently computed using the Matlab toolbox, i.e., CVX \cite{grant2015cvx1}, which only has a polynomial time complexity.

\subsection{Optimal Solution to ${\bf {P2-B}}$}

With the optimal $p_0^\star$ and  $B_0^\star$,  it is straightforward to know that problem $\bf {P2-B}$  is a convex optimization problem. The optimal solutions are given in the following theorem.
\begin{theorem}\label{theorem4v1}
The optimal solutions ${{{\bf{\tilde p}}^\star}}$ to problem $\bf {P2-B}$ is given by
\begin{align}
{{\tilde p}_{q,n}^\star} = {p_{q,n }^{\rm min}} + \max \left\{ {0,\frac{1}{{\ln 2\xi _n^\star}} - \frac{{1 + {\gamma _{q,n}}{p_{q,n }^{\rm min}}}}{{{\gamma _{q,n}}}}} \right\},\label{eqoptimalpower}
\end{align}
where ${p_{q,n }^{\rm min}} = \frac{1}{{{\gamma _{q,n}}}}\left( {{2^{\frac{M}{{{M_n'}}}{R_{q,\min }}}} - 1} \right)$ and  $\gamma_{q,n}$ is given in \eqref{eqrateZFMRT} with setting $p_0=p_0^\star$ and $B_0=B_0^\star$. Furthermore, ${{\xi_n^\star}}$ is the water level, which can be found by utilizing
enumeration and solving $\sum\nolimits_{q = 1}^Q {{{\tilde p}_{q,n}}}  = P$ for $1\le n \le N$.
\end{theorem}
\begin{IEEEproof}
This can be proved by using the standard Lagrange dual algorithm \cite{chen2018resource}, which is omitted here for brevity.
\end{IEEEproof}

Finally, the optimal solutions to problem ${\bf P2}$ can be obtained by finding the optimal $N$ that makes the objective maximum. The overall one-dimensional search based optimization algorithm is summarized in Algorithm 1

\subsection{Integer Conversion}
If we denote $({\bf p}^\dag,{\bf B}^\dag,\tilde{\bf  p}^\dag$,  $N^\dag)$  as the optimal solutions of problem ${\bf P2}$, the solution $ {\bf B}^\dag$ may violate the integer requirement, i.e., $\rm C10$, of the original problem ${\bf P1}$. Hence, we need to convert ${\bf B}^\dag$ into a feasible integer solution. However, considering the integer conversion problem is a combinatorial optimization problem that is challenging to solve optimally. Therefore, we develop the following heuristic algorithm to obtain its suboptimal solution. From \eqref{eq30t}, we know that a larger bandwidth will cause a better tracking performance. Thus, the integer solution, denoted by $B_k^\ddag$, should be larger than $B_k^\dag$ for $1\le k \le K$. Hence, we have the following  heuristic integer solution, i.e.,
 \begin{align}
B_k^\ddag  = \left\{ {\begin{array}{*{20}{c}}
{\left\lceil {B_k^\dag } \right\rceil ,\;{\rm{if}}\; 1 \le k \le K,}\\
{B - \sum\nolimits_{k = 1}^K {\left\lceil {B_k^\dag } \right\rceil ,\;{\rm{if }}\;k = 0} ,}
\end{array}} \right.\end{align}
where ${\left\lceil \cdot \right\rceil }$ is the ceiling function.

\begin{algorithm}
\caption{One-dimensional search based optimization algorithm for problem {\rm {\bf P}2}}\label{algorithmP0}
{\small{
\begin{algorithmic}[1]
 \STATE Calculate $N_{\rm max}$ using \eqref{eqNmaxa}
\FOR {channel aging time $N$=1:1:$N_{\rm max}$}
\STATE Solve ${\bf P2-A}$ and obtain $\left( {{{\bf{p}}^\star},{{\bf{B}}^\star}} \right)$

\STATE
With the optimal $p_0^\star$ and  $B_0^\star$ for the current $N$, solve ${\bf P2-B}$ and obtain ${\cal R}\left( {N,{{\bf{p}}^{\star}},{\bf{\tilde p}}^\star,{{\bf{B}}^{\star}}} \right)$
\ENDFOR
\STATE Find ${{\cal R}}\left( {N^\dag,{{\bf{p}}^{{\dag}}},{\bf{\tilde p}}^\dag,{{\bf{B}}^{\rm{\dag}}}} \right) = \mathop { \max }\limits_N {\cal R}\left(
 {N,{{\bf{p}}^{\star}},{\bf{\tilde p}}^\star,{{\bf{B}}^{\star}}} \right)$
\end{algorithmic}}}
\end{algorithm}

\begin{table*}[t]
 \centering
 %{\footnotesize
 \begin{tabular}{|c|c||c|c|}%{|l|p{5.2cm}||l|p{5.2cm}}
\hline
%{\textbf{Abbreviation}}& \textbf{Description}&{\textbf{Abbreviation}}& \textbf{Description}\\
%\hline
{Parameters}      &{Value}                                  &{Parameters}            &{Value}\\ \hline
{Speed of light}     &{$c_0$ = $3\times10^8$ m/s }                      &{Number of subcarriers} &{$B$ = 64}\\ \hline
{Total signal bandwidth} &{$B\Delta_f$ = 10 MHz }&{Subcarrier Bandwidth}&{$\Delta_f$ = 156.25 KHz} \\ \hline
 {Elementary OFDM symbol duration} &{$T_o$ = $1/\Delta_f$ = 6.4 us}  &{Cyclic prefix duration}      &{$T_{cp}$ = $\frac{1}{4}T$ = 1.6 us} \\ \hline
  Transmit OFDM symbol duration&{$T$ =  8 us}     &{Number of symbols }          &$M$ = 700\\ \hline
 \end{tabular}% }
\caption{Simulation parameter setup}\label{table1}
\end{table*}

\section{Simulation Results}

% The Doppler shift of the $q$-th communication receiver is given by $f_{q,D}=30 {\rm m/s}\times \frac{f_c}{c_0}$.

In this section, we show simulation results to validate the analyzed theoretical results and the effectiveness of the proposed algorithm.
Specifically,  we consider the system is operated on carrier frequency of $f_c=5.89$~GHz from IEEE 802.11p \cite{nguyen2017delay}.
The large-scale effect of communication channel is modeled by (74.2.4+16.1log$_{10}d_q^{\rm C}/d_o$) \cite{onubogu2014empirical}, where $d_o=1$m is the reference distance and $d_q^{\rm C}$ is the distance between the BS and communication receiver ${  U}_q^{\rm C}$, which follows continuous and uniformly distribution over $[1.5, 4.5]$~km. The temporal correlation coefficient ${\rho}_q$ is set as 0.96. %hemadeh2017millimeter
The noise power spectral densities at the BS and receiver $U_q^{\rm R}$ are set as -174 dBm/Hz.
%The receiver antenna gain at $U_q^{\rm R}$  is set as 20 dBi.
The initial angle, distance, and velocity of $U_k^{\rm R}$ follow continuous and uniformly distribution over $[\frac{(k-1)\pi}{16}$,$\frac{k\pi}{16}]$~rad, $[15, 45]$ m/s, and $[0.1, 0.3]$~km, respectively. Moreover, the associate state evaluation noise powers   are set as $\delta_k^\theta=10^{-5}$~rad, $\delta_k^d=0.2$ m, and $\delta_k^v=0.1$ m/s, respectively. Besides, the corresponding  maximum tracking errors are set as
 $\Gamma_{k,{\rm max}}^\theta=15 \delta_k^\theta$, $\Gamma_{k,{\rm max}}^d=15 \delta_k^d$, and $\Gamma_{k,{\rm max}}^v=15 \delta_k^v$.
We assume a uniform reflectivity model is adopted for each radar target and do not consider the specific RCSs for simplification  \cite{zhang2020power,godrich2011power}, i.e., ${\sigma _{{\rm{RCS,}}k}}$ = $1$ m$^2$.
The minimum transmission rate of $U_q^{\rm C}$ is set as $\frac{1}{10}{\log _2}\left( {1 + \frac{{P{\beta _q}}}{{{{\tilde \sigma }_q}Q}}} \right)$.
Finally, we use the similar setup in \cite{gaudio2019performance} for the remaining parameters, as shown in Table \ref{table1}.

In the following simulations, we compare the system performances obtained from our proposed algorithm and the following baseline schemes. Note that the results of each method are averaged over 1000 Monte Carlo trails.

\begin{figure*}
  \begin{minipage}{.325\textwidth}
   \centering
   \includegraphics[width=\textwidth]{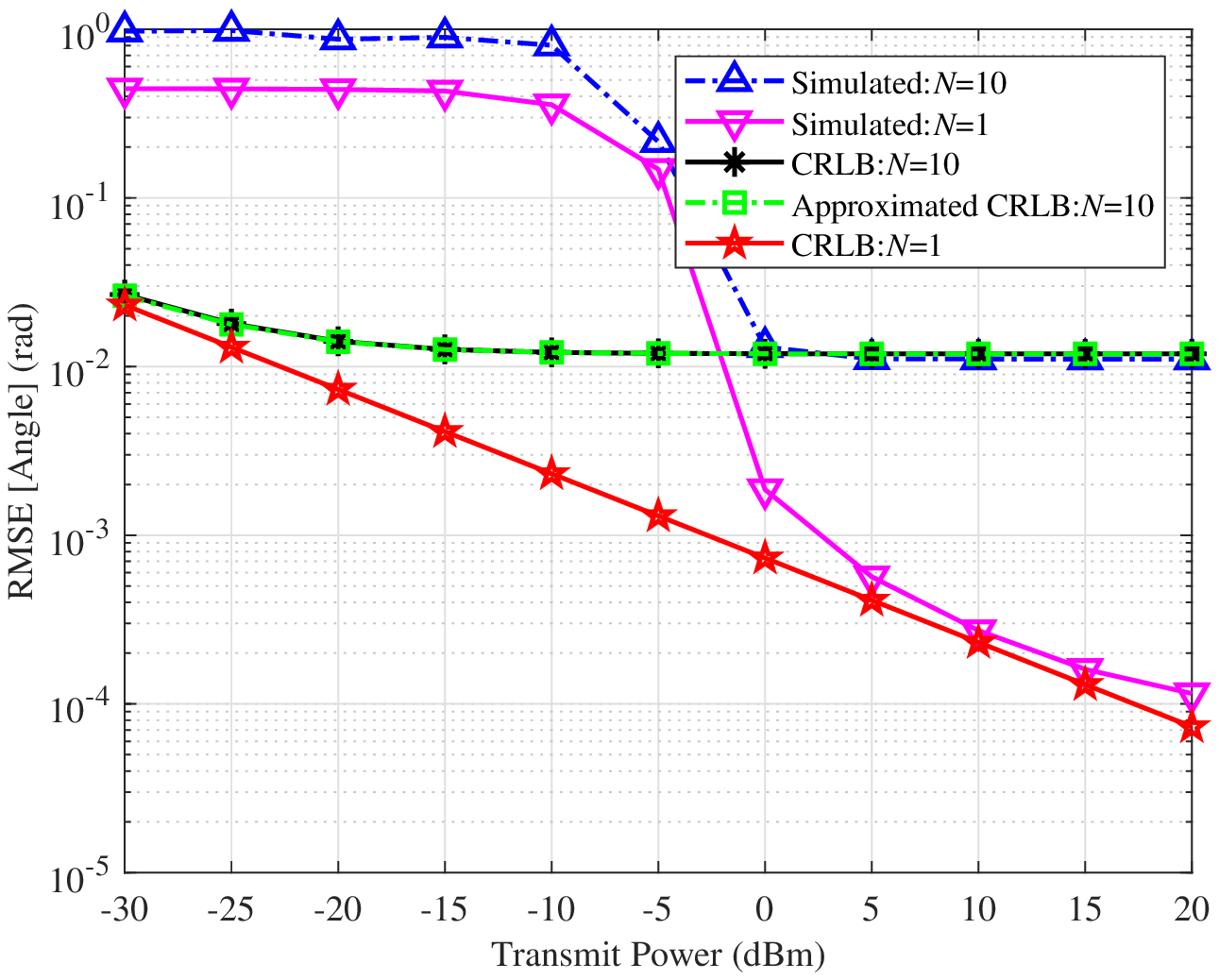}
   \caption*{\footnotesize (a) RMSE [Angle] }\label{figsimu11}
  \end{minipage}
 \begin{minipage}{.325\textwidth}
   \centering
   \includegraphics[width=\textwidth]{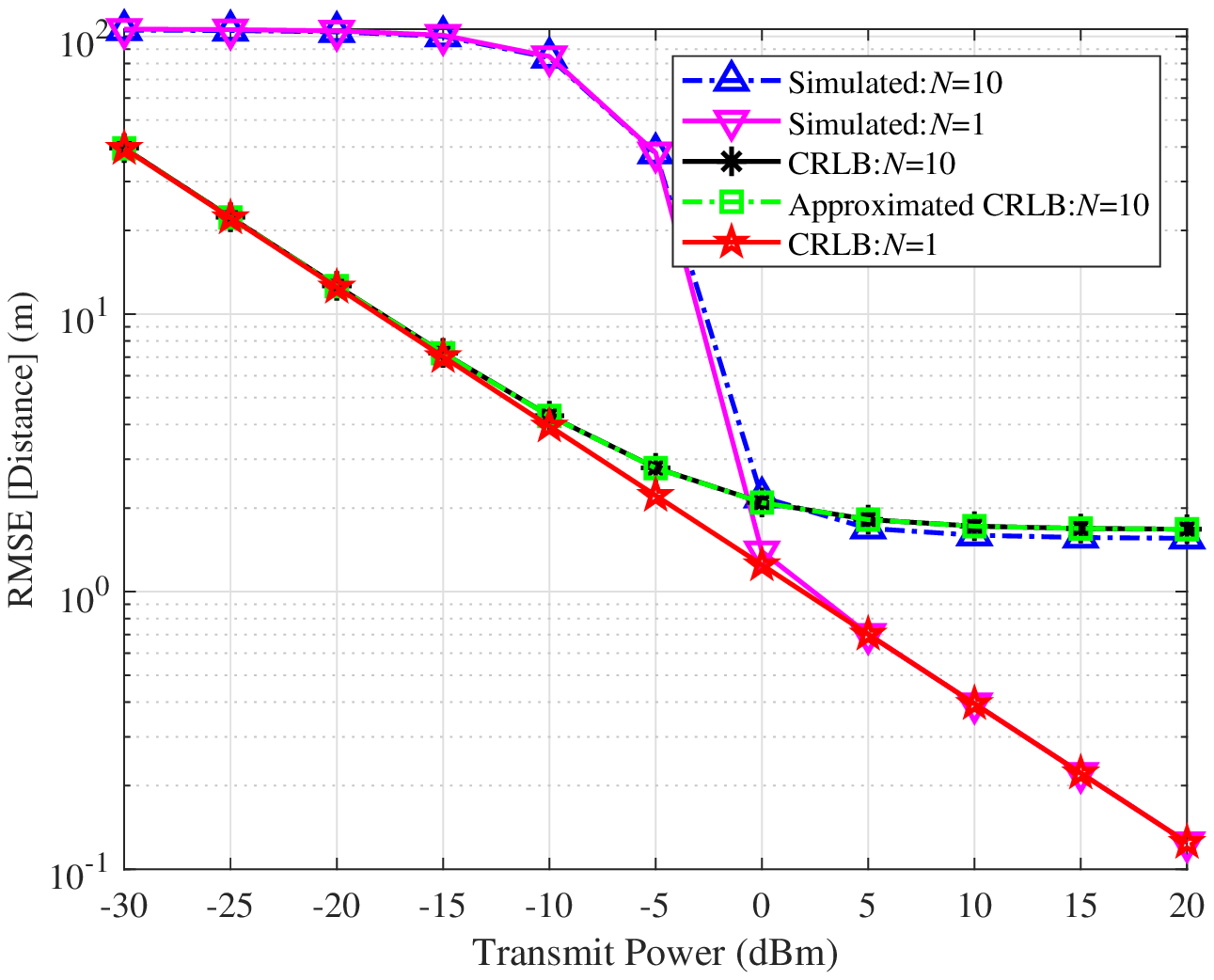}
    \caption*{\footnotesize (b)  RMSE [Distance]}\label{figsimu12}
  \end{minipage}
 \begin{minipage}{.325\textwidth}
   \centering
   \includegraphics[width=\textwidth]{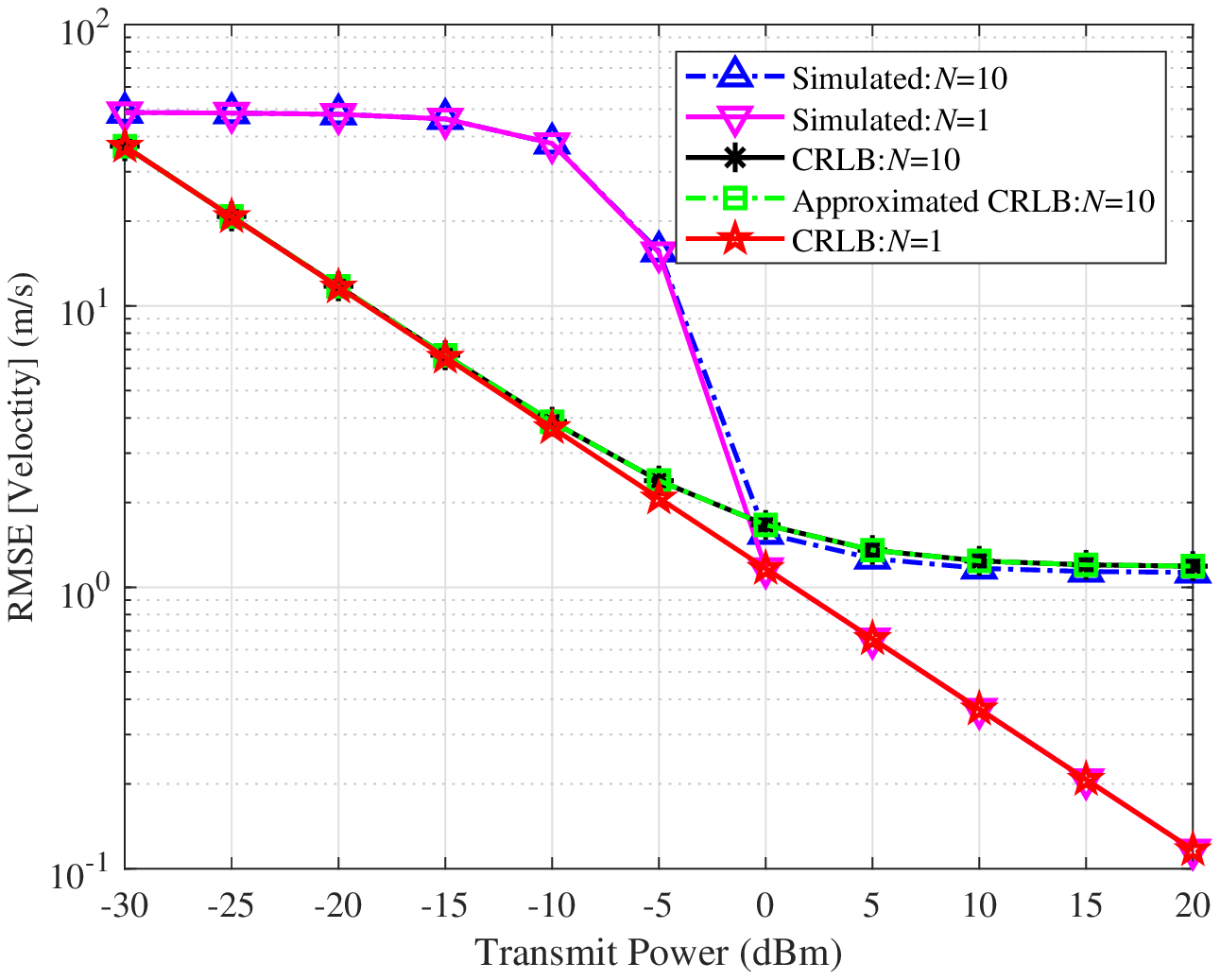}
    \caption*{\footnotesize (c) RMSE [Velocity]}\label{figsimu13}
  \end{minipage}
  \caption{The RMSEs versus the maximum transmit power with the uniform power and bandwidth allocation: $K$ = 4, $M_1$ = 60, $L_t=24$, and $L_r=64$. }\label{figsimu1}
 \end{figure*}

\begin{itemize}
\item Upper Bound: we obtain the optimal solution of the relaxed problem, ${\bf P2}$, using the proposed algorithm, without taking into account the integer constraint {\rm C10}. This relaxation allows us to obtain an upper bound that outperforms other algorithms.
\item Benchmark 1: we search for all potential $N$, and each is with optimized power and bandwidth by solving ${\bf P2-A}$ in Phase-I and uniform power in the remaining data transmission.
\item Benchmark 2: we search for all potential  $N$, and each is with uniform power and bandwidth allocation in Phase-I, but the optimized power allocating using \eqref{eqoptimalpower} in the remaining data transmission.
\item Benchmark 3:  we search for all potential $N$, and each is with uniform power and bandwidth for the entire transmission frame.
\item Without Channel Aging: we use the proposed algorithm to solve the original problem by setting $N=1$.
\end{itemize}

Fig. \ref{figsimu1} shows the impact of maximum transmit power on the simulated and theoretical radar tracking performances, i.e., the root of mean square error (RMSE), when the uniform power and bandwidth allocation is applied in the whole frame without considering sensing and communication performance requirements.
The RMSE is defined as $({\rm{RMSE}} = \sqrt {\frac{1}{K}\sum\nolimits_{k = 1}^K {{{\left( {{x_k} - {{\hat x}_k}} \right)}^2}} } )$, where $x_k$ and $\hat x_k$ are the true and estimated parameters, respectively.
Form Fig.~\ref{figsimu1}, we observe that the simulated RMSEs of the angle, distance, and velocity from \eqref{eqa112}, \eqref{eqa115}, and \eqref{eqa116}, can approach the theoretical results obtained by \eqref{eqcrlb} when $N=1$ and by \eqref{eq29t} when $N=10$, which validates the correctness of the analyzed radar channel estimation and prediction results.
Also, we observe that the proposed approximated results in \eqref{eq30t} can achieve nearly identical performances to the original theoretical results, which validates the effectiveness of the proposed approximation method.
Finally, we note that the tracking performances decrease as the channel aging time increases. This is because a larger channel aging time introduces higher evaluation noise in the radar channel model, which leads to worse tracking performance.
\begin{figure*}
  \begin{minipage}{.5\textwidth}
   \centering
    \includegraphics[width=\textwidth]{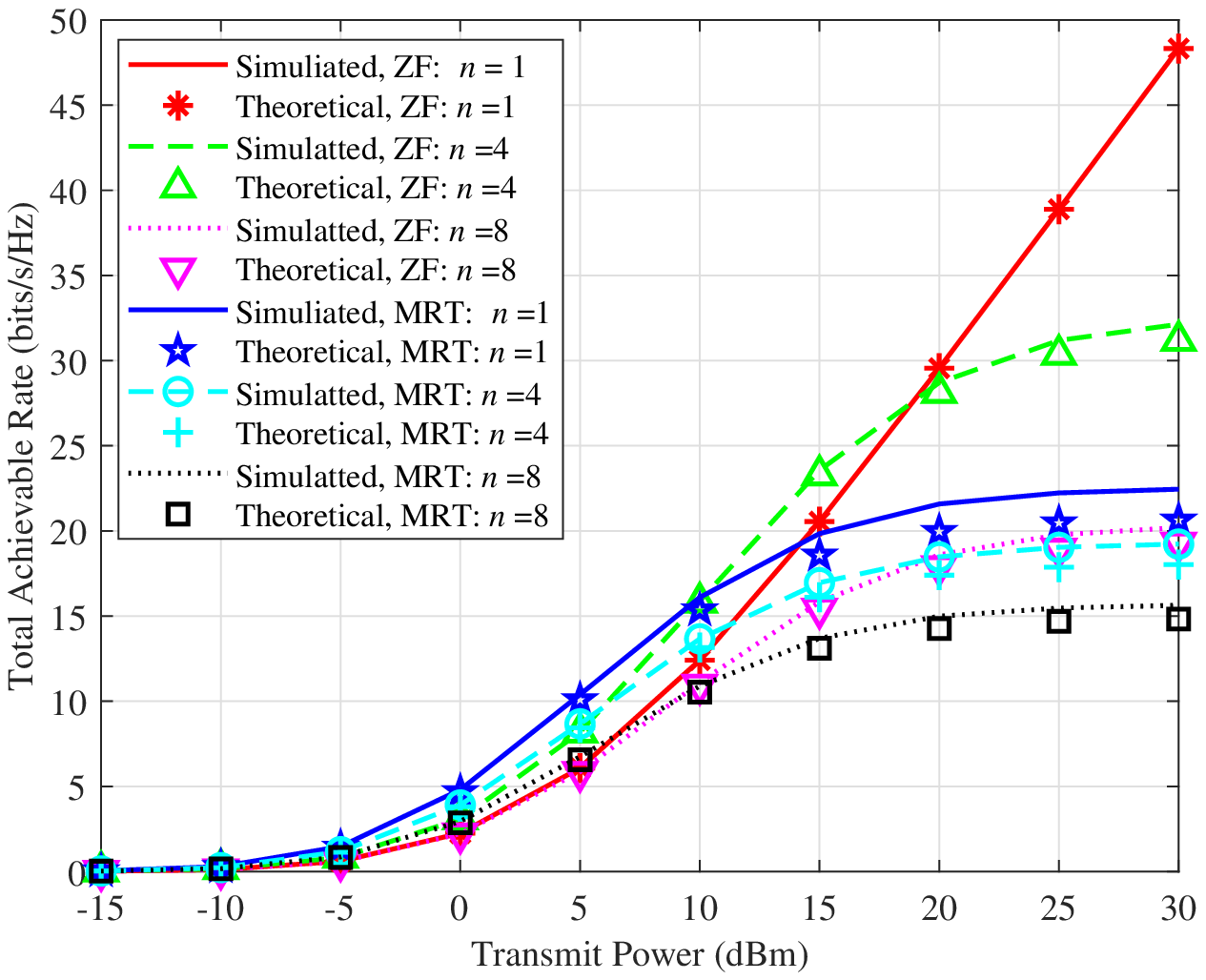}\vspace{-0.2cm}
   \caption*{\footnotesize (a) Simulated and theoretical results }\label{figsimu21}\vspace{-0.2cm}
  \end{minipage}
 \begin{minipage}{.5\textwidth}
   \centering
   \includegraphics[width=\textwidth]{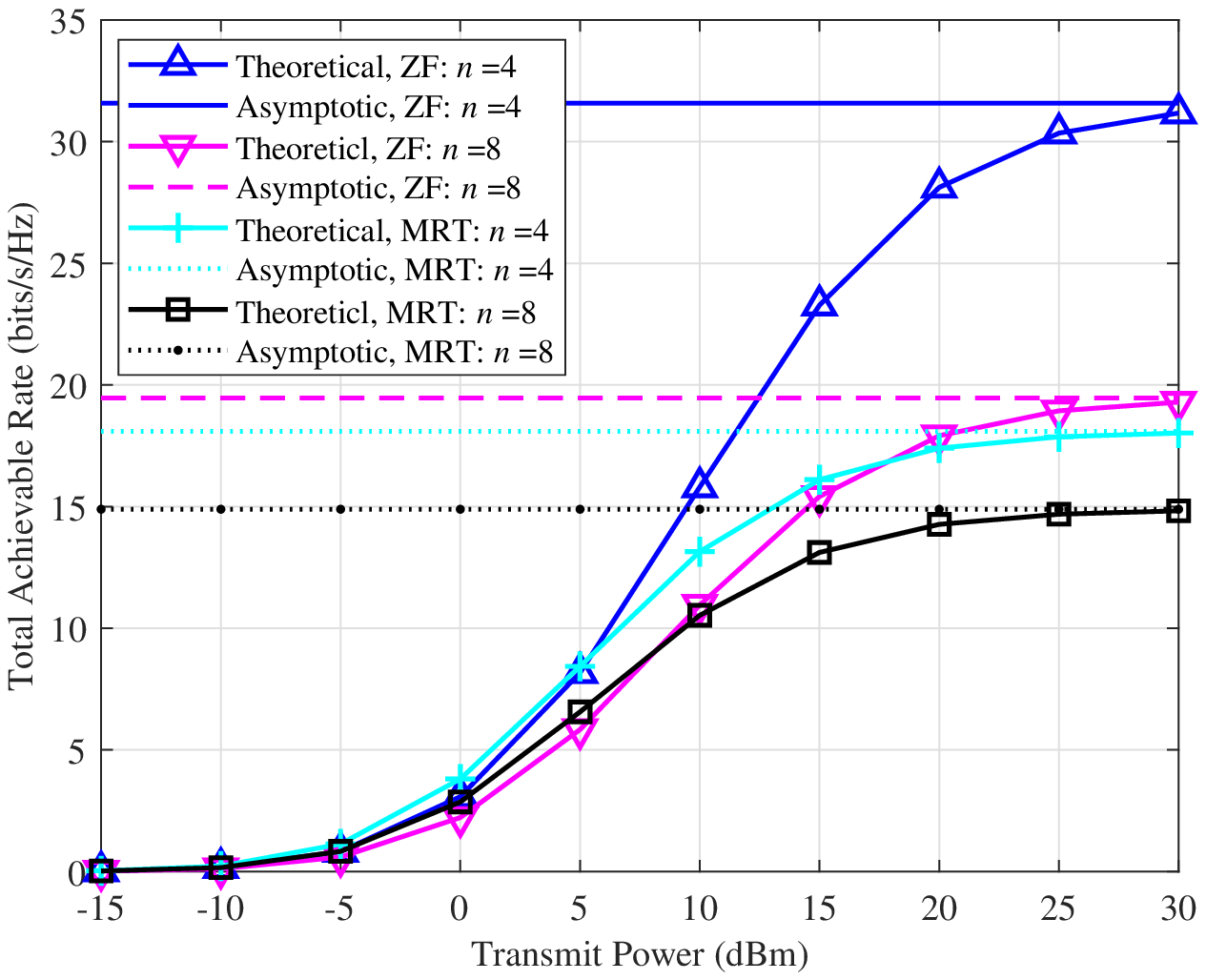}\vspace{-0.2cm}
    \caption*{\footnotesize (b)Asymptotic analysis}\label{figsimu22}\vspace{-0.2cm}
  \end{minipage}
  \caption{ The total achievable rate of all users in the $n$-th block versus the maximum transmit power with the uniform power and bandwidth allocation: $K$ = 4, $Q$ = 10, $M_1$ = 300, $L_t=32$, and $L_r=64$.}\label{figsimu2}\vspace{-0.2cm}
 \end{figure*}

%Fig. \ref{figsimu2} shows the impact of maximum transmit power on the simulated and theoretical communication performances, i.e., the
%total achievable transmission rate in Block $N$, i.e., ${R_n} = \sum\nolimits_{q = 1}^Q {{R_{q,n}}} $, when the uniform power and bandwidth allocation is applied in the whole transmission frame without considering sensing and communication performance requirements. From fig.~\ref{figsimu2}(a), we observe that the theoretical total achievable rates obtained from \eqref{eqrateZFMRT} match the simulated results, which verifies the correctness of the derived theoretical results.
%Also, we observe that these rates with MRT for any $N$ are bounded when the power increases. This is because there exists unavoidable beamforming interference in the MRT scheme with the non-orthogonal channels. However, the rates with ZF are not bounded only when $N=1$. The reason is that there is no interference in the ZF scheme, but this does not hold when $N\ge2$ due to the existence of channel evaluation noise.
%From fig.~\ref{figsimu2}(b), we observe that all results are bounded by the asymptotic results obtained in \eqref{eq47a}, which also validates the correctness of the derived theoretical results.

Fig. \ref{figsimu2} shows the impact of maximum transmit power on the simulated and theoretical communication performances.
Here we consider the total achievable transmission rate in the $n$-th block, denoted as ${R_n} = \sum\nolimits_{q = 1}^Q {{R_{q,n}}} $, when uniform power and bandwidth allocation is applied for the entire transmission frame, without considering sensing and communication performance requirements.
From Fig. \ref{figsimu2}(a), we observe that the theoretical total achievable rates obtained from \eqref{eqrateZFMRT} match the simulated results, which verifies the correctness of the derived theoretical results. Furthermore, we observe that the rates obtained using MRT for any $N$ are bounded as power increases due to the existence of interference in the MRT scheme with non-orthogonal channels.
In contrast, rates obtained using ZF are not bounded only when $N=1$, since no interference exists in the ZF scheme. However, this does not hold when $N\ge2$ due to the due to the existence of channel evaluation noise. From Fig. \ref{figsimu2}(b), we observe that all results are bounded by the asymptotic results obtained in \eqref{eq47a}, which verifies the derived theoretical results.

 \begin{figure}
  \begin{minipage}{.5\textwidth}
   \centering
   \includegraphics[width=\textwidth]{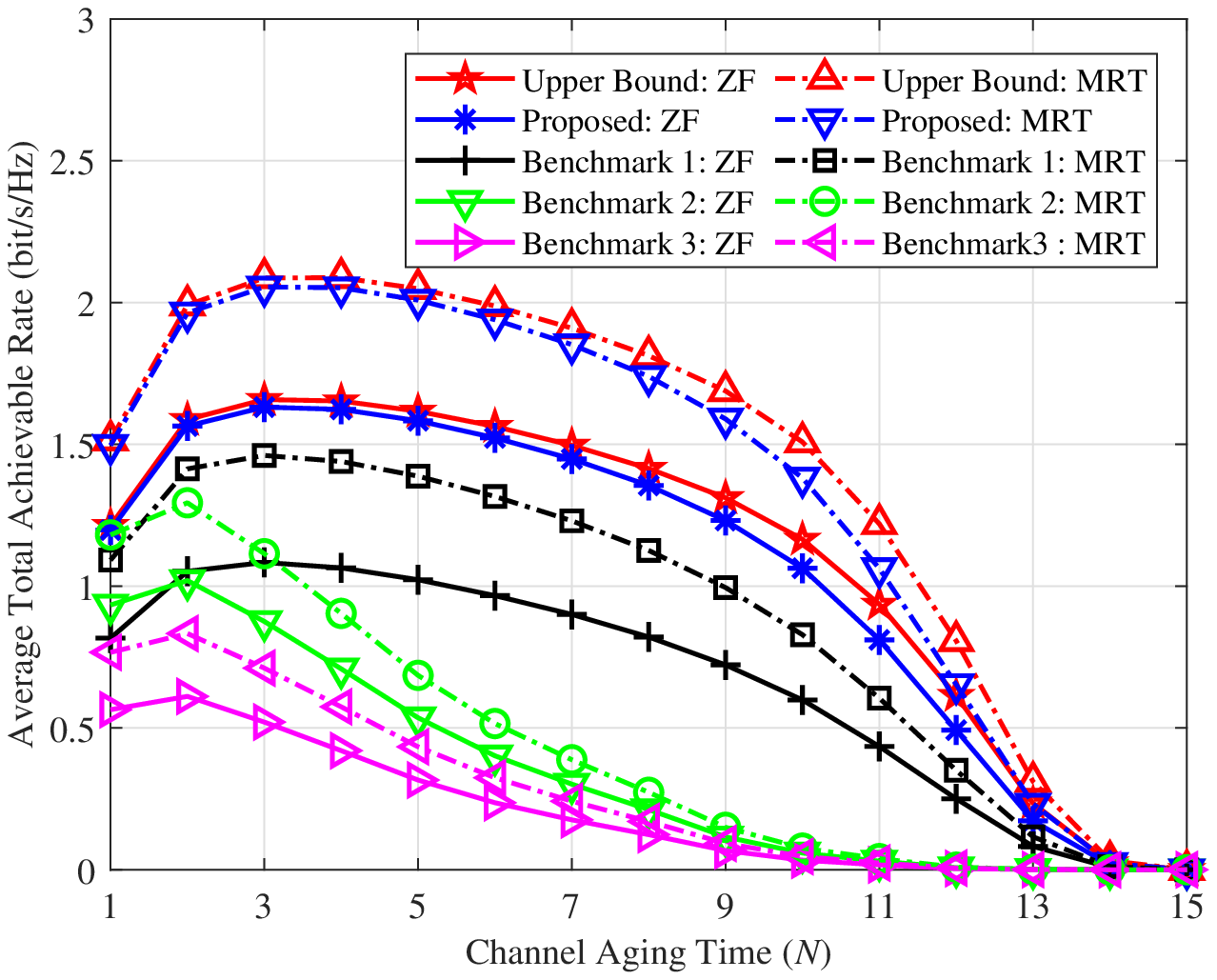}
   \caption{The average total achievable rate versus the channel aging time: $P$ = -5 dBm, $K$ = 4, $Q$ = 10, $M_1$ = 300, $L_t=32$, and $L_r=64$. }\label{figsimu3}\vspace{-0.4cm}
  \end{minipage}\quad
 \begin{minipage}{.5\textwidth}
   \centering
   \includegraphics[width=\textwidth]{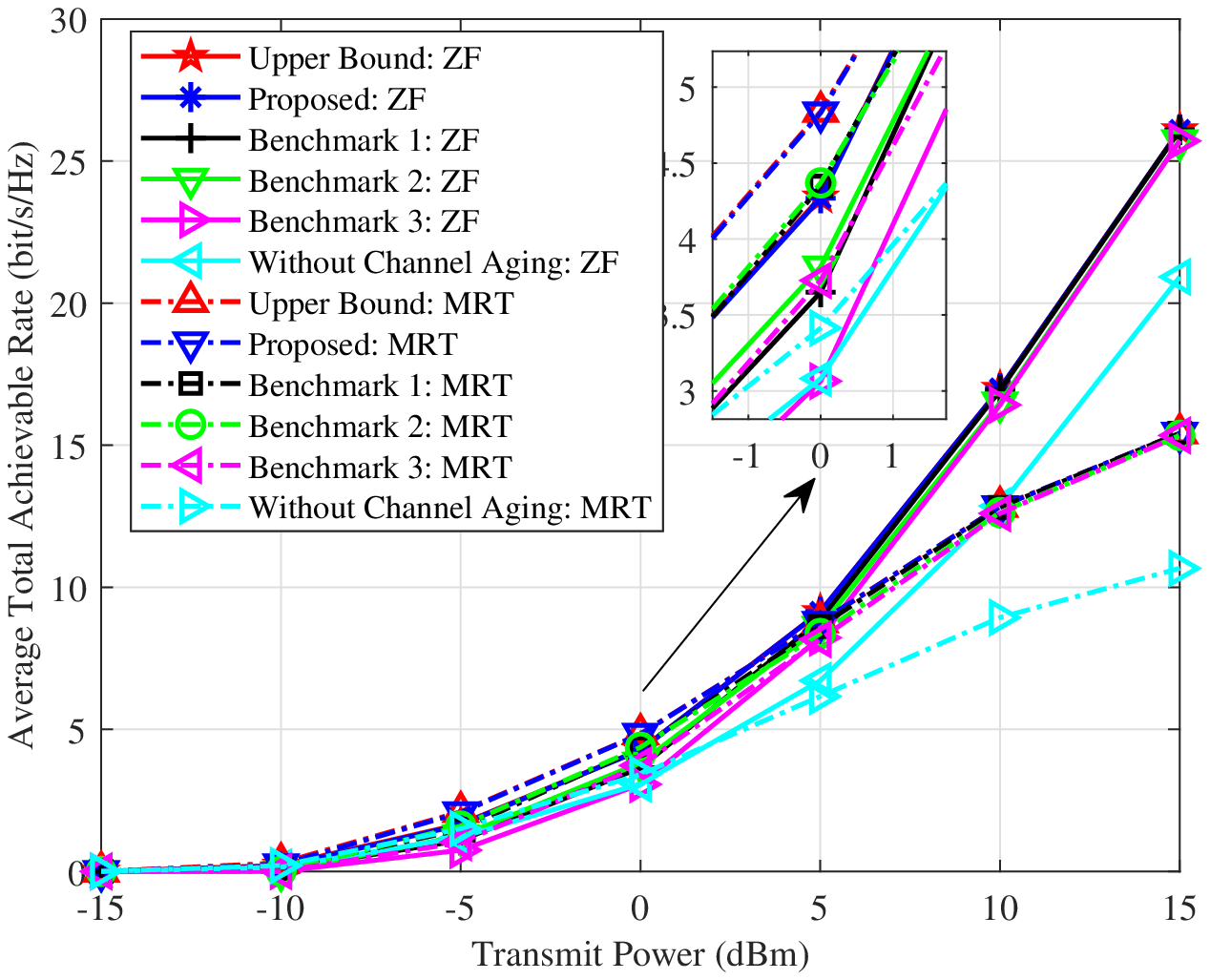}
   \caption{The average total achievable rate versus the maximum transmit power: $K$ = 4, $Q$ = 10, $M_1$ = 300, $L_t=32$, and $L_r=64$. }\label{figsimu4}\vspace{-0.4cm}
  \end{minipage}
 \end{figure}

Fig. \ref{figsimu3}  shows the impact of channel aging time on the average total achievable transmission rate. Note that in this figure, the channel aging time is a constant and does not need to be optimized in problem ${\bf P1}$. Hence, the proposed algorithm and other benchmarks with minor modifications can be applied to solve problem ${\bf P1}$ with a fixed $N$.
From this figure, we observe that the performance of each scheme initially increases and then decreases with the channel aging time. This is because exploiting the channel aging effect reduces the training overhead in CSI estimation and radar target tracking, which results in an increase in the achievable rate. However, the powers of the evaluation noises of the aged communication and radar channels also increase with the channel aging time, ultimately leading to a decrease in the rate. This observation motivates us to determine the optimal channel aging time that satisfies the specific system requirements and user demands.

Fig. \ref{figsimu4} shows the impact of maximum transmit power on the average total achievable transmission rate.
From this figure, we can observe that the total achievable rate of each scheme increases with the transmit power.
Besides, the proposed algorithm can achieve similar performance to the upper bound scheme and outperforms the baseline schemes, which demonstrates the effectiveness of the proposed algorithm.
Moreover, the performance gap between the proposed algorithm and the "Without Channel Aging" scheme increases with the transmit power.
The reason is that a better estimation/tracking performance can be obtained in the first block with a higher transmit power, thus causing a larger optimal channel aging time and higher transmission efficiency.
These findings validate  the importance of utilizing channel aging to improve system performance.
 Finally, it is observed that the algorithms with MRT beamforming outperform those with ZF beamforming in the low power region, but are inferior in the high power region. These results can guide the beamforming design for the studied system.

 \begin{figure}
  \begin{minipage}{.5\textwidth}
   \centering
   \includegraphics[width=\textwidth]{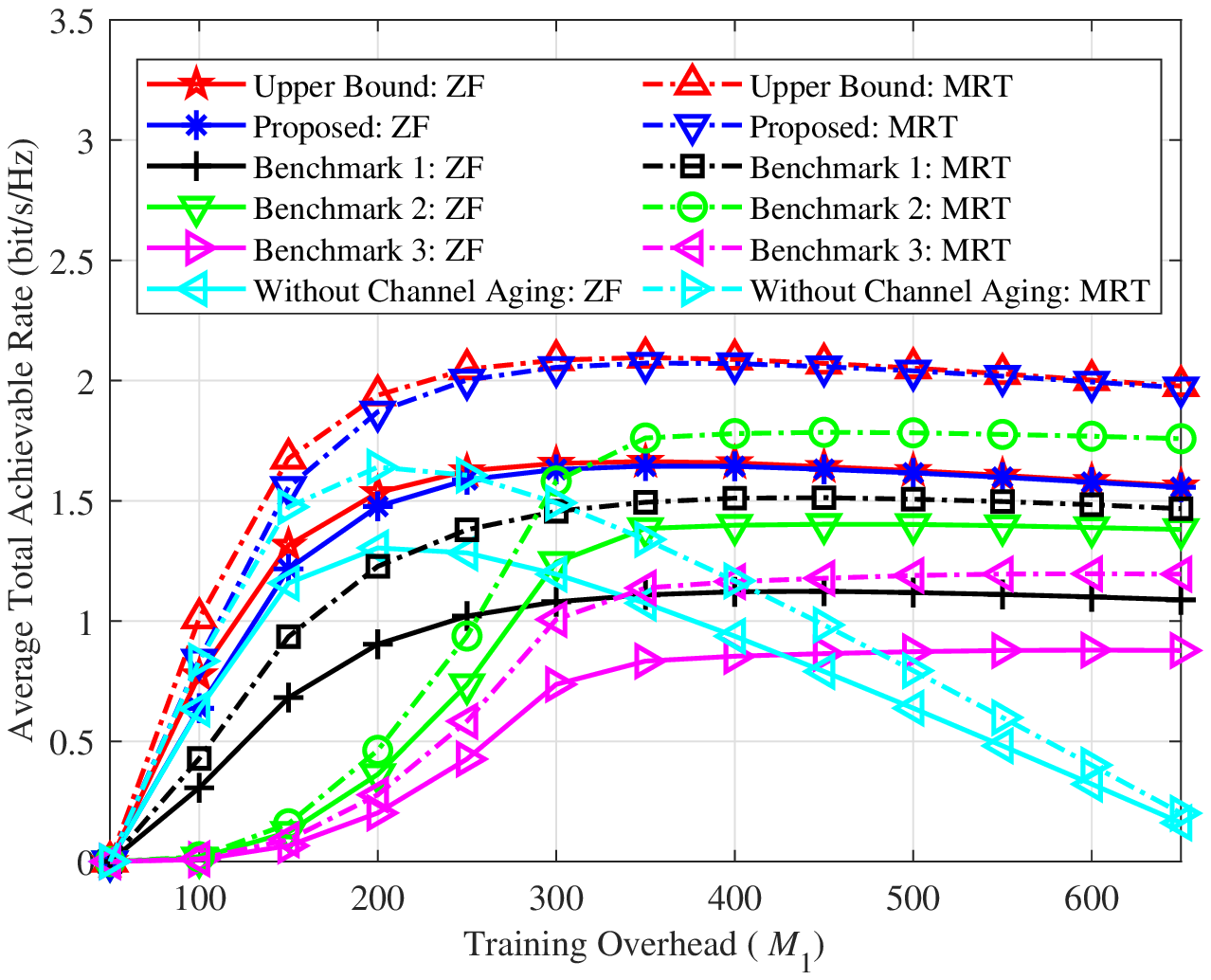}
   \caption{The average total achievable rate versus the training overhead: $P$ = -5 dBm, $K$ = 4, $Q$ = 10, $L_t=32$, and $L_r=64$.}\label{figsimu5}
  \end{minipage}\quad
 \begin{minipage}{.5\textwidth}
   \centering
   \includegraphics[width=\textwidth]{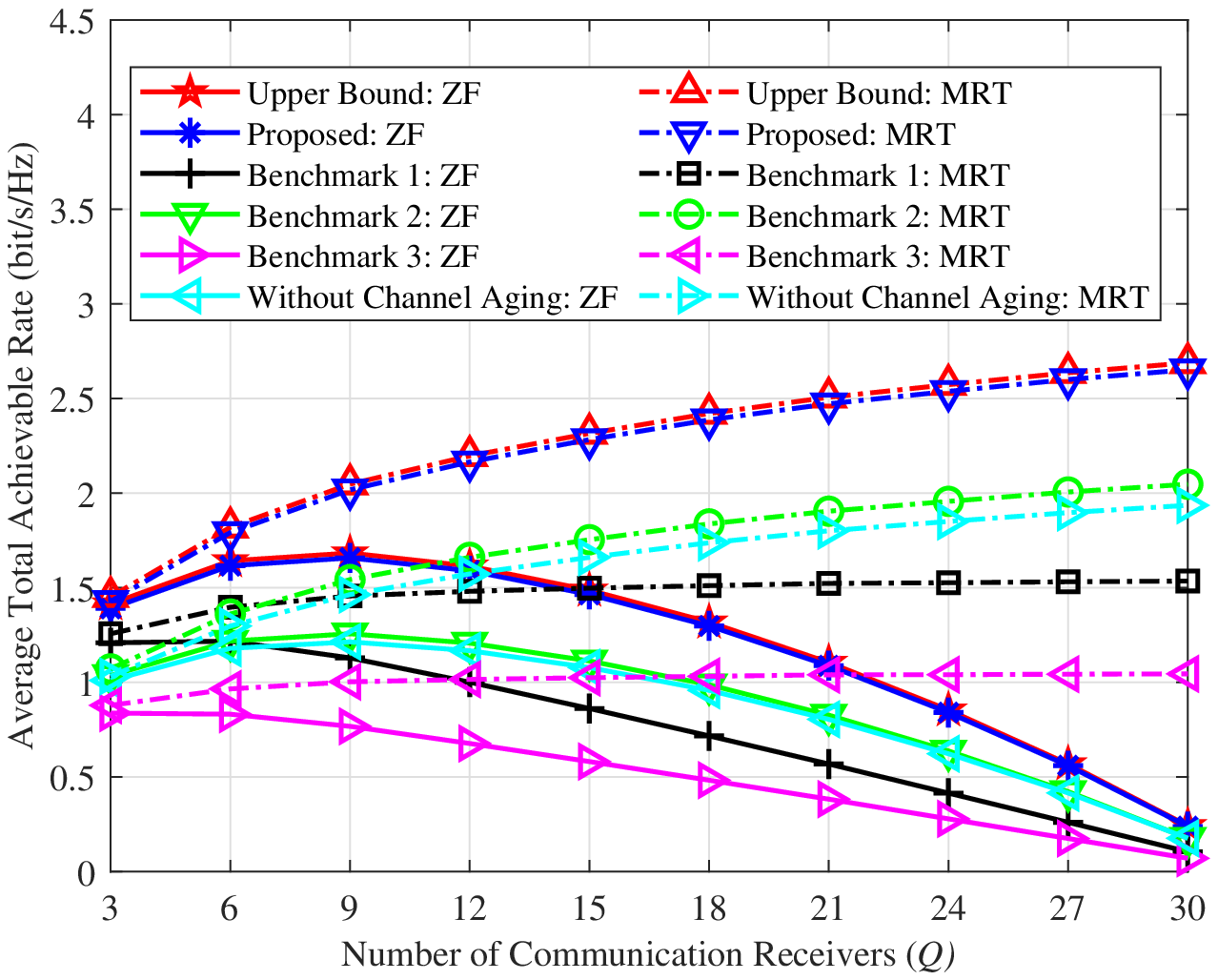}
   \caption{The average total achievable rate versus the number of communication receivers: $P$ = -5 dBm, $K$ = 4, $M_1$ = 300, $L_t=32$, and $L_r=64$.}\label{figsimu6}
  \end{minipage}
 \end{figure}

 Fig. \ref{figsimu5} shows the impact of training overhead on the average total achievable rate. It can be observed that the performance of each scheme increases initially and then decreases as the training overhead increases. The reason behind this is that a larger number of training signals can lead to better estimation/tracking performance, resulting in a higher achievable transmission rate. However, higher training overhead also implies less data transmission time in the first block, resulting in a lower achievable rate. Additionally, we can observe that the "Without Channel Aging" scheme's performance decreases faster than other schemes. This is because the training overhead exists in each block for this scheme, while it only exists in the first block for other schemes. This highlights the significant benefits of optimizing channel aging time in the system design.

Fig. \ref{figsimu6} shows the impact of the number of communication receivers on the average total achievable rate. From this figure, we observe that the performances of the MRT-based schemes increase with the number of users.
Because a larger number of users provides a larger degree of freedom for system design, thus causing a higher achievable rate.
However, the ZF-based schemes work worse with a larger number of users.
This is because ZF-based schemes need to guarantee orthogonality among all users. Thus, given the number of transmit antennas, the beamforming gain of the ZF-based scheme decreases with the number of users, thus causing a lower rate. This can be applied to guide the beamforming design for the studied system.

 \section{Conclusion}

This paper leverages channel aging characteristics to reduce the training overhead and proposes a situation-dependent channel re-estimation interval optimization-based resource allocation scheme for a downlink multi-user/-target DFRC system.
Specifically, we characterize the channel aging effects on the radar tracking CRLB performance and communication achievable rate performance, respectively. Then, these two performance metrics are derived as closed-form expressions with respect to the bandwidth, power, and channel aging time. Building on that, we formulate an average total aged achievable rate maximization problem subject to individual tracking precision demand,  customized communication rate requirement, and other practical constraints.
Moreover, the one-dimensional search based optimization algorithm is proposed to solve this problem efficiently.
Finally, simulation results validate the correctness of the derived results and the effectiveness of the proposed algorithm.

\appendix

\subsection{Proof of Theorem \ref{theorem0}}\label{appCRLB}
To simplify the notations when deriving CRLB, we ignore the subscript $k$ of
${{\bf {\overline y}}_k^{\rm R}}\left( {m,{b}} \right)$ in \eqref{eqMeasurement1} if there is no ambiguity and we denote the $(l+1)$-th entry of vector
${{\bf {\overline y}}_k^{\rm R}}\left( {m,{b}} \right)$ by ${y_{m,b,l}}$ for $0\le l\le L_r-1$. By denoting $\bar v=2\pi T\nu^D$ and $\bar \tau=2\pi\Delta_f \tau $ in \eqref{eqMeasurement1}, we have
 \begin{align}
{y_{m,b,l}} = \alpha {e^{j\left( {\phi  - l\pi sin(\theta)  + b\bar \tau + m\bar v} \right)}}{\rm{ + }}{\omega _{m,b,{l}}},\label{eqappendix1}
 \end{align}
where ${\omega _{m,b,{l }}}$ is the corresponding equivalent Gaussian noise with mean zero  and variance
$\frac{{ \sigma  }\Delta_f}{pL_t} $.

Then, by denoting ${\bf{z}} = {\left[ { \theta ,\bar \tau ,\bar v,\alpha ,\phi } \right]^T}$, we can obtain the corresponding fisher information matrix as
  \begin{scriptsize}
 \begin{align}
{{\bf{J}}_{ij}} = \frac{{2p{L_t}}}{{ \sigma {\Delta _f}}}\sum\limits_{m = 0}^{{M_1} - 1} {\sum\limits_{b = 0}^{{B_k} - 1} {\sum\limits_{l = 0}^{{L_r} - 1} {\left[ {\frac{{\partial y_{m,b,l}^{{\mathop{\rm Re}\nolimits} }}}{{\partial {z_i}}}\frac{{\partial y_{m,b,l}^{{\mathop{\rm Re}\nolimits} }}}{{\partial {z_j}}}{\rm{ + }}\frac{{\partial y_{m,b,l}^{{\mathop{\rm Im}\nolimits} }}}{{\partial {z_i}}}\frac{{\partial y_{m,b,l}^{{\mathop{\rm Im}\nolimits} }}}{{\partial {z_j}}}} \right]} } },\label{eqappendix2}
 \end{align}
 \end{scriptsize}
where $z_i$ is the $i$-th entry of vector ${\bf{z}} $ with $1\le i\le 5$, ${y_{m,b,l}^{{\rm{Re}}}}$ and $ {y_{m,b,l}^{{\rm{Im}}}}$ are the real and imaginary components of $y_{m,b,l}$, respectively. Then, the bounds of the estimation error covariance are
 \begin{align}
{\mathop{\rm var}} \left( {{{\hat z}_i}} \right) \ge {{\bf{J}}^{ - 1}}\left( {i,i} \right)\label{eqappendix3},
 \end{align}
where ${{\bf{J}}^{ - 1}}\left( {i,i} \right)$ is the $i$-th diagonal element of the inverse matrix $\bf J$.

Since it is quite difficult to derive the closed-form bounds from \eqref{eqappendix3} by using all the measurements due to the complexity of inverse processing, we use the similar method in \cite{braun2014ofdm} to obtain a simpler lower bound. For calculating the bound of ${\theta}$, we can use the fact that each measurement has a different, random, and unknown initial phase due to the unknown Doppler shift and time delay. Since the white noise is assumed as the source of error, the estimation of $\theta$ from each measurement can be regarded as independent \cite{braun2014ofdm}. By applying the probability theory, we have
 \begin{align}
{\mathop{ \rm var}} \left( {\hat { \theta} } \right) \ge \frac{1}{{{M_1}{B_k}}}{\bf{J}}_{\theta }^{ - 1}\left( {1,1} \right),\label{eqappendix4}
 \end{align}
where
${\bf{J}}_{{\theta }} = \frac{{2p{L_t}}}{{ \sigma {\Delta _f}}}\sum\nolimits_{l = 0}^{{L_r} - 1} {\left[ {\frac{{\partial y_{m,b,l}^{{\mathop{\rm Re}\nolimits} }}}{{\partial {z_i}}}\frac{{\partial y_{m,b,l}^{{\mathop{\rm Re}\nolimits} }}}{{\partial {z_j}}}{\rm{ + }}\frac{{\partial y_{m,b,l}^{{\mathop{\rm Im}\nolimits} }}}{{\partial {z_i}}}\frac{{\partial y_{m,b,l}^{{\mathop{\rm Im}\nolimits} }}}{{\partial {z_j}}}} \right]} ,i,j \in \left\{ {1,4,5} \right\}$. Substituting \eqref{eqappendix1} into \eqref{eqappendix4}, we have
 \begin{align}
{\mathop{\rm var}} \left( {\hat {\theta} }\right) \ge\frac{{6 \sigma\Delta_f }}{{{p}\left| {{\alpha }} \right|^2{\pi ^2}{{\cos }^2}\left( \theta  \right){B_k}{M_1}{L_tL_r}\left( {L_r^2 - 1} \right)}}.
\end{align}
Similarly, we can define ${\bf{J}}_{{\bar \tau }} = \frac{{2p{L_t}}}{{ \sigma {\Delta _f}}}\sum\nolimits_{b = 0}^{{B_k} - 1} {\left[ {\frac{{\partial y_{m,b,l}^{{\mathop{\rm Re}\nolimits} }}}{{\partial {z_i}}}\frac{{\partial y_{m,b,l}^{{\mathop{\rm Re}\nolimits} }}}{{\partial {z_j}}}{\rm{ + }}\frac{{\partial y_{m,b,l}^{{\mathop{\rm Im}\nolimits} }}}{{\partial {z_i}}}\frac{{\partial y_{m,b,l}^{{\mathop{\rm Im}\nolimits} }}}{{\partial {z_j}}}} \right]} ,i,j \in \left\{ {2,4,5} \right\}$ and  ${\bf{J}}_{{\bar v }} = \frac{{2p{L_t}}}{{ \sigma {\Delta _f}}}\sum\nolimits_{m = 0}^{{M_1} - 1} {\left[ {\frac{{\partial y_{m,b,l}^{{\mathop{\rm Re}\nolimits} }}}{{\partial {z_i}}}\frac{{\partial y_{m,b,l}^{{\mathop{\rm Re}\nolimits} }}}{{\partial {z_j}}}{\rm{ + }}\frac{{\partial y_{m,b,l}^{{\mathop{\rm Im}\nolimits} }}}{{\partial {z_i}}}\frac{{\partial y_{m,b,l}^{{\mathop{\rm Im}\nolimits} }}}{{\partial {z_j}}}} \right]} ,i,j \in \left\{ {3,4,5} \right\}$ to derive the bounds of $\bar \tau$ and $\bar v$, respectively, and then we have
 \begin{align}
{\mathop{\rm var}} \left( {\hat {\bar \tau} }\right) \ge\frac{{6 \sigma \Delta_f}}{{{p}\left| {{\alpha }} \right|^2 {B_k}{M_1}{L_tL_r}\left( {B_k^2 - 1} \right)}},\label{eqaooendix66}\\
{\mathop{\rm var}} \left( {\hat {\bar v} }\right) \ge\frac{{6 \sigma\Delta_f }}{{{p}\left| {{\alpha }} \right|^2 {B_k}{M_1}{L_tL_r}\left( {M_1^2 - 1} \right)}}\label{eqaooendix67}.
\end{align}
Then, since $\bar \tau=2\pi\Delta_f \tau $ and ${\tau } = 2\frac{{{d}}}{{{c_0}}}$, we have $\bar \tau=4\pi\Delta_f \frac{{{d}}}{{{c_0}}} $. Also, since $\bar v=2\pi T\nu^D$ and $\nu^D = \frac{{2{v}}\cos \left( {{\theta} - \varphi_k } \right)}{c_0}{f_c}$, we have $\bar v=4\pi T{f_c}\frac{{ {v}}\cos \left( {{\theta} - \varphi_k } \right)}{c_0}$. Finally, based on the bounds derived in \eqref{eqaooendix66} and \eqref{eqaooendix67},  we can obtain the CRLBs of distance $d$ and velocity $v$ as
  \begin{footnotesize} \begin{align}
&{\mathop{\rm var}} \left( {\hat { d} }\right) \ge\frac{{3c_0^2 \sigma\Delta_f }}{{8{{\left( {\pi \Delta_f} \right)}^2}p\left| \alpha  \right|^2B_k{M_1}{L_tL_r}\left( {{B_k^2} - 1} \right)}},\\
&{\mathop{\rm var}} \left( {\hat { v} }\right) \ge\frac{{3c_0^2 \sigma\Delta_f}}{{8{{\left( {\pi {T }} \right)}^2}f_c^2p\left| \alpha  \right|^2 {{\cos }^2}\left( {\theta  - \varphi_k } \right)B_k{M_1}{L_tL_r}\left( {M_1^2 - 1} \right)}}.
\end{align}
 \end{footnotesize}

\subsection{Proof of Theorem \ref{theorem3}}\label{appendixAppr1}

To simplify the notations when deriving the approximated CRLBs, we ignore the subscript $k$ of all variables if there is no ambiguity.
To begin with, the Jacobian matrix for ${\partial {\cal G}\left( {{{\bf{x}}}} \right)}$ is defined as $ {{{\bf{G}}_{{n}}} = \frac{{\partial {\cal G}\left( {{{\bf{x}}}} \right)}}{{\partial {{\bf{x}}}}}\left| {_{{{\bf{x}}} = {{{\bf{\hat x}}_{n}}}}} \right.}$. With \eqref{eqARCI1} and \eqref{eqARCI2}, the Jacobian matrix ${{\bf{G}}_{{n}}}$ can be calculated as
\begin{small}
\begin{align}
{{\bf{G}}_n}=\left[ {\begin{array}{*{20}{c}}
{1{\rm{ + }}\frac{{{{\hat v}_n} \widetilde T}}{{{{\hat d}_n}}}\cos \left( {{{{\hat {\tilde \theta}} }_n}} \right)}&{ - \frac{{{{\hat v}_n} \widetilde T}}{{\hat d_n^2}}\sin \left( {{{ {\hat {\tilde \theta}} }_n}} \right)}&{\frac{{  \widetilde T}}{{{{\hat d}_n}}}\sin \left( {{{\hat {\tilde \theta}}_n}} \right)}\\
{{{\hat v}_n} \widetilde T\sin \left( {{{\hat {\tilde \theta}}_n}} \right)}&1&{ -  \widetilde T\cos \left( {{{{\hat {\tilde \theta}} }_n}} \right)}\\
0&0&1
\end{array}} \right],
\end{align}
\end{small}
where ${{\hat {\tilde \theta}} _n} = {\hat \theta _n} - {\varphi _k}$.
Since  $\frac{{{{\hat v}_n}{{\hat v}_{n - 1}} {{\widetilde T}^2}}}{{{{\hat d}_n}{{\hat d}_{n - 1}}}} \approx 0$, $\frac{{{{\hat v}_n}\widetilde T}}{{\hat d_n^2}} \approx 0$, and $\frac{{{{\hat v}_n}{{\hat v}_{n - 1}}{\widetilde T^2}}}{{{{\hat d}_n}}} \approx 0$,  ${{\bf{G}}_n}{{\bf{G}}_{n-1}}$ can be approximated by
\begin{small}
\begin{align}
{{\bf{G}}_n}{{\bf{G}}_{n - 1}} \approx \left[ {\begin{array}{*{20}{c}}
{1{\rm{ + }}\sum\limits_{i = n - 1}^n {\frac{{{{\hat v}_i} \widetilde T}}{{{{\hat d}_i}}}\cos \left( {{{{\hat {\tilde \theta}} }_i}} \right)} }&0&{\sum\limits_{i = n - 1}^n {\frac{{   \widetilde T}}{{{{\hat d}_i}}}\sin \left( {{{{\hat {\tilde \theta}} }_i}} \right)} }\\
{\sum\limits_{i = n - 1}^n {{{\hat v}_i}  \widetilde T\sin \left( {{{{\hat {\tilde \theta}} }_i}} \right)} }&1&{ - \sum\limits_{i = n - 1}^n { \widetilde T\cos \left( {{{{\hat {\tilde \theta}} }_i}} \right)} }\\
0&0&1
\end{array}} \right]\label{eqappedix35}.
\end{align}
\end{small}

Then, by some algebraic manipulations with utilizing \eqref{eqappedix35}, we have
\begin{small}
\begin{align}
\prod\limits_{i' = i}^{n - 1} {{{\bf{G}}_{i'}}}  \approx \left[ {\begin{array}{*{20}{c}}
{1{\rm{ + }}\sum\limits_{i' = i}^{n - 1} {\frac{{{{\hat v}_{i'}} \widetilde T}}{{{{\hat d}_{i'}}}}\cos \left( {{{{\hat {\tilde \theta}} }_{i'}}} \right)} }&0&{\sum\limits_{i' = i}^{n - 1} {\frac{{  \widetilde T}}{{{{\hat d}_{i'}}}}\sin \left( {{{{\hat {\tilde \theta}} }_{i'}}} \right)} }\\
{\sum\limits_{i' = i}^{n - 1} {{{\hat v}_{i'}}  \widetilde T\sin \left( {{{{\hat {\tilde \theta}} }_{i'}}} \right)} }&1&{ - \sum\limits_{i' = i}^{n - 1} {  \widetilde T\cos \left( {{{{\hat {\tilde \theta}} }_{i'}}} \right)} }\\
0&0&1
\end{array}} \right].
\end{align}
\end{small}
Similarly, due to $\frac{{{{\hat v}_i}{{\hat v}_{i' }} {\widetilde T^2}}}{{\hat d}_i{{{\hat d}_{i'}}}} \approx 0$ for $i \ne i'$,
the first and second terms of the right hand side of \eqref{eq28b} can be approximated as the following two equations, respectively, i.e.,
\begin{footnotesize}\begin{align}
 {\rm{diag}}\left( {{{{\bf{\tilde G}}}_{n - 1}}{\bf{D\tilde G}}_{n - 1}^H} \right) &\approx {\rm{diag}}\left( {{a_n},1,1} \right) {\bf{D}} \label{eqappendix38a},\\
 {\rm{diag}}\left( {\sum\limits_{i = 1}^{n - 1} {{{{\bf{\bar G}}}_{n,i}}} {{\bf{\Sigma }}_\omega }{\bf{\bar G}}_{n,i}^H} \right) &\approx \left( {n - 1} \right){\rm{diag}}\left( {\frac{{{b_n}}}{{n - 1}},1,1} \right){{\bf{\Sigma }}_\omega }
 \label{eqappendix39a},
\end{align}
\end{footnotesize}
where $a_n  = {\left| {1 + \sum\nolimits_{i = 1}^{n - 1} {\frac{{{{\hat v}_i} \widetilde T}}{{{{\hat d}_i}}}\cos \left( {{{{\hat {\tilde \theta}} }_i}} \right)} } \right|^2}$ and ${b_n} = 1 + \sum\nolimits_{i = 2}^{n - 1} {{{\left| {1 + \sum\nolimits_{i' = i}^{n - 1} {\frac{{{{\hat v}_{i'}}  \widetilde T}}{{{{\hat d}_{i'}}}}\cos \left( {{{{\hat {\tilde \theta}} }_{i'}}} \right)} } \right|}^2}} $.

Finally, substituting \eqref{eqappendix38a} and \eqref{eqappendix39a} into \eqref{eq28b}, the CRLBs derived in \eqref{eq29t} can be approximated as \eqref{eq30t} and the proof is completed.

\subsection{Proof of Theorem \ref{Theorem31}}\label{appendixTheorem31}
 When the MRT beamforming   is applied, we have ${{\bf{f}}_{q,n}} = \frac{1}{{\sqrt {{\lambda _q}{L_t}} }}{{{\bf{\hat h}}}_{q,1}}$. Then, we have the following terms:
\begin{footnotesize}
\begin{align}
&{\left| {{\mathbb E}\left( {{\bf{h}}_{q,n}^H{{\bf{f}}_{q,n}}} \right)} \right|^2} \nonumber \\
=&{\left| {{\mathbb E}\left( {\left( {{\bf{\hat h}}_{q,n}^H + {\bf{\bar e}}_{q,n}^H} \right)\frac{1}{{\sqrt {{\lambda _q}{L_t}} }}{{{\bf{\hat h}}}_{q,1}}} \right)} \right|^2}
= {\left| {{\mathbb E}\left( {\frac{{{\rho_q ^{n - 1}}}}{{\sqrt {{\lambda _q}{L_t}} }}{\bf{\hat h}}_{q,1}^H{{{\bf{\hat h}}}_{q,1}}} \right)} \right|^2}\nonumber \\
=&{\rho_q ^{2\left( {n - 1} \right)}}{\lambda _q}{L_t} \label{eqzfappend1},\\
&{\mathbb E}\left( {{{\left| {{\bf{h}}_{q,n}^H{{\bf{f}}_{q,n}} - {\mathbb E}\left( {{\bf{h}}_{q,n}^H{{\bf{f}}_{q,n}}} \right)} \right|}^2}} \right)\nonumber \\
=  & {\mathbb E}\left( {{{\left| {\frac{{{\rho_q ^{n - 1}}}}{{\sqrt {{\lambda _q}{L_t}} }}{\bf{\hat h}}_{q,1}^H{{{\bf{\hat h}}}_{q,1}}} \right|}^2}} \right) \!+\! {\mathbb E}\left( {{{\left| {\frac{1}{{\sqrt {{\lambda _q}{L_t}} }}{\bf{\bar e}}_{q,n}^H{{{\bf{\hat h}}}_{q,1}}} \right|}^2}} \right) \!-\! {\left| {{\mathbb E}\left( {{\bf{h}}_{q,n}^H{{\bf{f}}_{q,n}}} \right)} \right|^2}\nonumber\\
 = &{\rho_q ^{2\left( {n - 1} \right)}}{\lambda _q}\left( {{L_t} + 1} \right) + \left( {{\beta _q} - {\rho_q ^{2\left( {n - 1} \right)}}{\lambda _q}} \right) - {\rho_q ^{2\left( {n - 1} \right)}}{\lambda _q}{L_t} = {\beta _q}\label{eqzfappend2},\\
&{\mathbb E}\left( {{{\left| {\left( {{\bf{\hat h}}_{q,n}^H + {\bf{\bar e}}_{q,n}^H} \right){{\bf{f}}_{i,n}}} \right|}^2}} \right)\nonumber \\
 = &{\mathbb E}\left( {{{\left| {\frac{{{\rho_q ^{n - 1}}}}{{\sqrt {{\lambda _i}{L_t}} }}{\bf{\hat h}}_{q,1}^H{{{\bf{\hat h}}}_{i,1}}} \right|}^2}} \right) + {\mathbb E}\left( {{{\left| {\frac{1}{{\sqrt {{\lambda _i}{L_t}} }}{\bf{\bar e}}_{q,n}^H{{{\bf{\hat h}}}_{i,1}}} \right|}^2}} \right)\nonumber\\
 =& {\rho_q ^{2\left( {n - 1} \right)}}{\lambda _q} + \left( {{\beta _q} - {\rho_q ^{2\left( {n - 1} \right)}}{\lambda _q}} \right)
 =  {\beta _q}\label{eqzfappend3}.
\end{align}
\end{footnotesize}
Then,  by substituting \eqref{eqzfappend1}, \eqref{eqzfappend2}, and \eqref{eqzfappend3} into \eqref{eqsinr}, the   channel-to-interference-plus-noise ratio $\gamma_{q,n}$ under MRT is given in \eqref{eqrateZFMRT}.

When the ZF beamforming   is applied, we have ${{\bf{f}}_{q,n}} =\sqrt {{\lambda _q}\left( {{L_t} - Q} \right)} {{\bf{a}}_q}$.
According to \cite{ngo2013energy}, ${\bf{\hat h}}_{q,1}^H{{\bf a}_i} =1$ if $q=i$, and 0 otherwise. Then, we can derive the following terms:
\begin{footnotesize}\begin{align}
&{\left| {{\mathbb E}\left( {{\bf{h}}_{q,n}^H{{\bf{f}}_{q,n}}} \right)} \right|^2}\nonumber \\
 =& {\left| {{\mathbb E}\left( {\left( {{\bf{\hat h}}_{q,n}^H + {\bf{\bar e}}_{q,n}^H} \right)\sqrt {{\lambda _q}\left( {{L_t} - Q} \right)} {{\bf{a}}_q}} \right)} \right|^2} {\rm{ = }}{\rho_q ^{2\left( {n - 1} \right)}}{\lambda _q}\left( {{L_t} - Q} \right) \label{eqzfappend4},\\
&{\mathbb E}\left( {{{\left| {{\bf{h}}_{q,n}^H{{\bf{f}}_n} - {\mathbb E}\left( {{\bf{h}}_{q,n}^H{{\bf{f}}_n}} \right)} \right|}^2}} \right)\nonumber \\
 = &  {\mathbb E}\left( {{{\left| {\left( {{\rho_q ^{n - 1}}{\bf{\hat h}}_{q,1}^H + {\bf{\bar e}}_{q,n}^H} \right)\sqrt {{\lambda _q}\left( {{L_t} -Q} \right)} {{\bf{a}}_q}} \right|}^2}} \right) - {\left| {{\mathbb E}\left( {{\bf{h}}_{q,n}^H{{\bf{f}}_{q,n}}} \right)} \right|^2}\nonumber\\
 = &{\lambda _q}\left( {{L_t} - Q} \right)\left( {{\beta _q} - {\rho_q ^{2\left( {n - 1} \right)}}{\lambda _q}} \right){\mathbb E}\left( {{{\left\| {{{\bf{a}}_q}} \right\|}^2}} \right)
 =  {{\beta _q} - {\rho_q ^{2\left( {n - 1} \right)}}{\lambda _q}}  \label{eqzfappend5},\\
&{\mathbb E}\left( {{{\left| {\left( {{\bf{\hat h}}_{q,n}^H + {\bf{\bar e}}_{q,n}^H} \right){{\bf{f}}_{i,n}}} \right|}^2}} \right) \nonumber \\
=& {\mathbb E}\left( {{{\left| {{\rho_q ^{n - 1}}{\bf{\hat h}}_{q,1}^H\sqrt {{\lambda _i}\left( {{L_t} - Q} \right)} {{\bf{a}}_i}} \right|}^2}} \right) + {\mathbb E}\left( {{{\left| {{\bf{\bar e}}_{q,n}^H\sqrt {{\lambda _i}\left( {{L_t} - Q} \right)} {{\bf{a}}_i}} \right|}^2}} \right)\nonumber\\
 =&  {{\beta _q} - {\rho_q ^{2\left( {n - 1} \right)}}{\lambda _q}}  \label{eqzfappend6}.
\end{align}
\end{footnotesize}
Finally,  by substituting  \eqref{eqzfappend4}, \eqref{eqzfappend5}, and \eqref{eqzfappend6} into \eqref{eqsinr}, the   channel-to-interference-plus-noise ratio $\gamma_{q,n}$ under ZF is given in \eqref{eqrateZFMRT}.
Finally, the proof is completed.

\subsection{Proof of Theorem \ref{theoremtransform}}\label{theoremtransformapp}

For notation simplicity in the following proof, the objective function ${\cal R}\left( {N,{\bf{p}},{\bf{\tilde p}},{\bf{B}}} \right)$ with fixed $N$ can be expressed as the function ${\rm{F}}\left( {{p_0}{B_{0}},{\bf{\tilde p}}} \right)$, which is only dependent on ${p_0}{B_{0}}$ and ${\bf{\tilde p}}$. In the following, we apply the contradiction method to prove the transform equivalence.

Firstly, since problem $\bf {P2}$ with the fixed $N$ is feasible, we assume that
${\bf{p}}^\star = \left[ {{p_0^\star},{p_1^\star},\cdots,{p_K^\star}} \right]^{T} $ and
${\bf{B}}^\star = \left[ {{B_0^\star},{B_1^\star},\cdots,{B_K^\star}} \right]^{T} $ are the corresponding optimal solutions. Also, we assume that
${\bf{p}}^\dag = \left[ {{p_0^\dag},{p_1^\dag},\cdots,{p_K^\dag}} \right]^{T}$ and
${\bf{B}}^\dag = \left[ {{B_0^\dag},{B_1^\dag},\cdots,{B_K^\dag}} \right]^{T} $ are the optimal solutions  to  problem ${\bf P2-A}$ with the same fixed $N$.  Based on the above assumptions, we know ${\bf{p}}^\dag$, ${\bf{B}}^\dag$, and ${\bf{\tilde p}}^*$  are also the feasible  solutions of problem $\bf {P2}$ with the fixed $N$.
Besides, we have the following important results based on the above assumptions, i.e., ${\rm{F}}\left( {{p_0^\star}{B_0^\star},{\bf{\tilde p}}^\star} \right)\ge{\rm{F}}\left( {{p_0^\dag}{B_0{^\dag}},{\bf{\tilde p}}^\star} \right)$ and $p_0^\star B_0^\star\leq p_0^\dag B_0^\dag$. However, it is straightforward to know that  ${\rm{F}}\left( {{p_0}{B_{0}},{\bf{\tilde p}}} \right)$ is a monotonic increasing function with respect to $p_0B_0$. Thus, if $p_0^\star B_0^\star\leq p_0^\dag B_0^\dag$, we have ${\rm{F}}\left( {{p_0^\star}{B_0^\star},{\bf{\tilde p}}^\star} \right)\le{\rm{F}}\left( {{p_0^\dag}{B_0{^\dag}},{\bf{\tilde p}}^\star} \right)$. This contradicts with the assumption that $p_0^\star$ and $B_0^\star$  are the optimal solutions to problem $\bf {P2}$ with the fixed $N$. Hence, we know problems $\bf {P2}$ and $\bf {P2}-A$ have the same solutions of ${\bf{p}}$ and ${\bf{B}}$. Recalling that ${\rm{F}}\left( {{p_0}{B_{0}},{\bf{\tilde p}}} \right)$  is only dependent on ${p_0}{B_{0}}$ and ${\bf{\tilde p}}$. Thus, once $p_0B_0$ is obtained, it is straightforward to know that  the optimal ${\bf{\tilde p}}$ of problem $\bf {P2}$  can be obtained by solving problem $\bf {P2}-B$. Hence, the theorem is proved.

\ifCLASSOPTIONcaptionsoff
  \newpage
\fi
  \bibliography{SPT}

% Generated by IEEEtran.bst, version: 1.14 (2015/08/26)
\begin{thebibliography}{10}
\providecommand{\url}[1]{#1}
\csname url@samestyle\endcsname
\providecommand{\newblock}{\relax}
\providecommand{\bibinfo}[2]{#2}
\providecommand{\BIBentrySTDinterwordspacing}{\spaceskip=0pt\relax}
\providecommand{\BIBentryALTinterwordstretchfactor}{4}
\providecommand{\BIBentryALTinterwordspacing}{\spaceskip=\fontdimen2\font plus
\BIBentryALTinterwordstretchfactor\fontdimen3\font minus
  \fontdimen4\font\relax}
\providecommand{\BIBforeignlanguage}[2]{{%
\expandafter\ifx\csname l@#1\endcsname\relax
\typeout{** WARNING: IEEEtran.bst: No hyphenation pattern has been}%
\typeout{** loaded for the language `#1'. Using the pattern for}%
\typeout{** the default language instead.}%
\else
\language=\csname l@#1\endcsname
\fi
#2}}
\providecommand{\BIBdecl}{\relax}
\BIBdecl

\bibitem{zheng2019radar}
L.~Zheng, M.~Lops, Y.~C. Eldar, and X.~Wang, ``Radar and communication
  coexistence: An overview: A review of recent methods,'' \emph{IEEE Signal
  Process. Mag.}, vol.~36, no.~5, pp. 85--99, 2019.

\bibitem{liu2022integrated}
F.~Liu, Y.~Cui, C.~Masouros, J.~Xu, T.~X. Han, Y.~C. Eldar, and S.~Buzzi,
  ``Integrated sensing and communications: Towards dual-functional wireless
  networks for 6{G} and beyond,'' \emph{IEEE J. Sel. Areas Commun.}, vol.~40,
  no.~6, pp. 1728--1767, 2022.

\bibitem{sturm2011waveform}
C.~Sturm and W.~Wiesbeck, ``Waveform design and signal processing aspects for
  fusion of wireless communications and radar sensing,'' \emph{Proc. IEEE},
  vol.~99, no.~7, pp. 1236--1259, 2011.

\bibitem{liu2020joint}
F.~Liu, C.~Masouros, A.~P. Petropulu, H.~Griffiths, and L.~Hanzo, ``Joint radar
  and communication design: Applications, state-of-the-art, and the road
  ahead,'' \emph{IEEE Trans. Commun.}, vol.~68, no.~6, pp. 3834--3862, 2020.

\bibitem{yang2020queue}
H.~Yang, Z.~Wei, Z.~Feng, C.~Qiu, Z.~Fang, X.~Chen, and P.~Zhang, ``Queue-aware
  dynamic resource allocation for the joint communication-radar system,''
  \emph{IEEE Trans. Veh. Technol.}, vol.~70, no.~1, pp. 754--767, 2020.

\bibitem{ouyang2022noma}
C.~Ouyang, Y.~Liu, and H.~Yang, ``{NOMA}-{ISAC}: Performance analysis and rate
  region characterization,'' \emph{arXiv preprint arXiv:2205.13756}, 2022.

\bibitem{tian2022adaptive}
T.~Tian, G.~Li, H.~Deng, and J.~Lu, ``Adaptive bit/power allocation with
  beamforming for dual-function radar-communication,'' \emph{IEEE Wireless
  Commun. Lett.}, vol.~11, no.~6, pp. 1186--1190, Jun. 2022.

\bibitem{temiz2021optimized}
M.~Temiz, E.~Alsusa, and M.~W. Baidas, ``Optimized precoders for massive {MIMO}
  {OFDM} dual radar-communication systems,'' \emph{IEEE Trans. Commun.},
  vol.~69, no.~7, pp. 4781--4794, 2021.

\bibitem{luo2019optimization}
Y.~Luo, J.~A. Zhang, X.~Huang, W.~Ni, and J.~Pan, ``Optimization and
  quantization of multibeam beamforming vector for joint communication and
  radio sensing,'' \emph{IEEE Trans. Commun.}, vol.~67, no.~9, pp. 6468--6482,
  2019.

\bibitem{zhou2021performance}
X.~Zhou, L.~Tang, Y.~Bai, and Y.-C. Liang, ``Performance analysis and waveform
  optimization of integrated {FD}-{MIMO} radar-communication systems,''
  \emph{IEEE Trans. Wireless Commun.}, vol.~20, no.~11, pp. 7490--7502, 2021.

\bibitem{wang2020constrained}
X.~Wang, Z.~Fei, J.~A. Zhang, J.~Huang, and J.~Yuan, ``Constrained utility
  maximization in dual-functional radar-communication multi-{UAV} networks,''
  \emph{IEEE Trans. Commun.}, vol.~69, no.~4, pp. 2660--2672, 2020.

\bibitem{wang2022noma}
Z.~Wang, Y.~Liu, X.~Mu, and Z.~Ding, ``{NOMA} inspired interference
  cancellation for integrated sensing and communication,'' in \emph{Proc. IEEE
  Inter. Conf. Commun. (ICC)}, 2022, pp. 3154--3159.

\bibitem{ashraf2022detection}
M.~Ashraf and B.~Tan, ``Detection probability maximization scheme in integrated
  sensing and communication systems,'' in \emph{Proc. IEEE Veh. Technol. Conf.
  (VTC-Spring)}, 2022, pp. 1--6.

\bibitem{cui2013mimo}
G.~Cui, H.~Li, and M.~Rangaswamy, ``{MIMO} radar waveform design with constant
  modulus and similarity constraints,'' \emph{IEEE Trans. Signal Process.},
  vol.~62, no.~2, pp. 343--353, 2013.

\bibitem{hua2022integrated}
M.~Hua, Q.~Wu, W.~Chen, and A.~Jamalipour, ``Integrated sensing and
  communication: Joint pilot and transmission design,'' \emph{arXiv preprint
  arXiv:2211.12891}, 2022.

\bibitem{cao2020joint}
N.~Cao, Y.~Chen, X.~Gu, and W.~Feng, ``Joint radar-communication waveform
  designs using signals from multiplexed users,'' \emph{IEEE Trans. Commun.},
  vol.~68, no.~8, pp. 5216--5227, 2020.

\bibitem{ni2021multi}
Z.~Ni, J.~A. Zhang, K.~Yang, X.~Huang, and T.~A. Tsiftsis, ``Multi-metric
  waveform optimization for multiple-input single-output joint communication
  and radar sensing,'' \emph{IEEE Trans. Commun.}, vol.~70, no.~2, pp.
  1276--1289, 2021.

\bibitem{xie2017joint}
M.~Xie, W.~Yi, T.~Kirubarajan, and L.~Kong, ``Joint node selection and power
  allocation strategy for multitarget tracking in decentralized radar
  networks,'' \emph{IEEE Trans. Signal Process.}, vol.~66, no.~3, pp. 729--743,
  2017.

\bibitem{yuan2020spatio}
X.~Yuan, Z.~Feng, J.~A. Zhang, W.~Ni, R.~P. Liu, Z.~Wei, and C.~Xu,
  ``Spatio-temporal power optimization for {MIMO} joint communication and radio
  sensing systems with training overhead,'' \emph{IEEE Trans. Veh. Technol.},
  vol.~70, no.~1, pp. 514--528, 2020.

\bibitem{kumari2019adaptive}
P.~Kumari, S.~A. Vorobyov, and R.~W. Heath, ``Adaptive virtual waveform design
  for millimeter-wave joint communication--radar,'' \emph{IEEE Trans. Signal
  Process.}, vol.~68, pp. 715--730, 2019.

\bibitem{9913311}
M.~Hua, Q.~Wu, C.~He, S.~Ma, and W.~Chen, ``Joint active and passive
  beamforming design for {IRS}-aided radar-communication,'' \emph{IEEE Trans.
  Wireless Commun.}, pp. 1--1, 2022, doi: 10.1109/TWC.2022.3210532.

\bibitem{yan2015simultaneous}
J.~Yan, H.~Liu, B.~Jiu, B.~Chen, Z.~Liu, and Z.~Bao, ``Simultaneous multibeam
  resource allocation scheme for multiple target tracking,'' \emph{IEEE Trans.
  Signal Process.}, vol.~63, no.~12, pp. 3110--3122, 2015.

\bibitem{yan2014power}
J.~Yan, H.~Liu, B.~Jiu, and Z.~Bao, ``Power allocation algorithm for target
  tracking in unmodulated continuous wave radar network,'' \emph{IEEE Sens.
  J.}, vol.~15, no.~2, pp. 1098--1108, 2014.

\bibitem{zhang2020power}
H.~Zhang, B.~Zong, and J.~Xie, ``Power and bandwidth allocation for
  multi-target tracking in collocated {MIMO} radar,'' \emph{IEEE Trans. Veh.
  Technol.}, vol.~69, no.~9, pp. 9795--9806, 2020.

\bibitem{muns2019beam}
G.~R. Muns, K.~V. Mishra, C.~B. Guerra, Y.~C. Eldar, and K.~R. Chowdhury,
  ``Beam alignment and tracking for autonomous vehicular communication using
  {IEEE} 802.11 ad-based radar,'' in \emph{Proc. IEEE Conf. Comput. Commun.
  Workshops (INFOCOMWKSHPS)}, 2019, pp. 535--540.

\bibitem{yuan2020bayesian}
W.~Yuan, F.~Liu, C.~Masouros, J.~Yuan, D.~W.~K. Ng, and
  N.~Gonz{\'a}lez-Prelcic, ``Bayesian predictive beamforming for vehicular
  networks: A low-overhead joint radar-communication approach,'' \emph{IEEE
  Trans. Wireless Commun.}, vol.~20, no.~3, pp. 1442--1456, 2020.

\bibitem{liu2020radar}
F.~Liu, W.~Yuan, C.~Masouros, and J.~Yuan, ``Radar-assisted predictive
  beamforming for vehicular links: Communication served by sensing,''
  \emph{IEEE Trans. Wireless Commun.}, vol.~19, no.~11, pp. 7704--7719, 2020.

\bibitem{liu2022learning}
C.~Liu, W.~Yuan, S.~Li, X.~Liu, H.~Li, D.~W.~K. Ng, and Y.~Li, ``Learning-based
  predictive beamforming for integrated sensing and communication in vehicular
  networks,'' \emph{IEEE J. Sel. Areas Commun.}, vol.~40, no.~8, pp.
  2317--2334, 2022.

\bibitem{zhou1999tracking}
Y.~Zhou, P.~C. Yip, and H.~Leung, ``Tracking the direction-of-arrival of
  multiple moving targets by passive arrays: Algorithm,'' \emph{IEEE Trans.
  Signal Process.}, vol.~47, no.~10, pp. 2655--2666, 1999.

\bibitem{chen2018waveform}
P.~Chen, C.~Qi, L.~Wu, and X.~Wang, ``Waveform design for kalman filter-based
  target scattering coefficient estimation in adaptive radar system,''
  \emph{IEEE Trans. Veh. Technol.}, vol.~67, no.~12, pp. 11\,805--11\,817,
  2018.

\bibitem{baddour2005autoregressive}
K.~E. Baddour and N.~C. Beaulieu, ``Autoregressive modeling for fading channel
  simulation,'' \emph{IEEE Trans. Wireless Commun.}, vol.~4, no.~4, pp.
  1650--1662, 2005.

\bibitem{chopra2016throughput}
R.~Chopra, C.~R. Murthy, and H.~A. Suraweera, ``On the throughput of large
  {MIMO} beamforming systems with channel aging,'' \emph{IEEE Signal Process.
  Lett.}, vol.~23, no.~11, pp. 1523--1527, 2016.

\bibitem{kong2015sum}
C.~Kong, C.~Zhong, A.~K. Papazafeiropoulos, M.~Matthaiou, and Z.~Zhang,
  ``Sum-rate and power scaling of massive {MIMO} systems with channel aging,''
  \emph{IEEE Trans. Commun.}, vol.~63, no.~12, pp. 4879--4893, 2015.

\bibitem{papazafeiropoulos2016impact}
A.~K. Papazafeiropoulos, ``Impact of general channel aging conditions on the
  downlink performance of massive {MIMO},'' \emph{IEEE Trans. Veh. Technol.},
  vol.~66, no.~2, pp. 1428--1442, Feb. 2016.

\bibitem{deng2019intermittent}
R.~Deng, Z.~Jiang, S.~Zhou, and Z.~Niu, ``Intermittent {CSI} update for massive
  {MIMO} systems with heterogeneous user mobility,'' \emph{IEEE Trans.
  Commun.}, vol.~67, no.~7, pp. 4811--4824, 2019.

\bibitem{zheng2021impact}
J.~Zheng, J.~Zhang, E.~Bj{\"o}rnson, and B.~Ai, ``Impact of channel aging on
  cell-free massive {MIMO} over spatially correlated channels,'' \emph{IEEE
  Trans. Wireless Commun.}, vol.~20, no.~10, pp. 6451--6466, 2021.

\bibitem{gaudio2019performance}
L.~Gaudio, M.~Kobayashi, B.~Bissinger, and G.~Caire, ``Performance analysis of
  joint radar and communication using {OFDM} and {OTFS},'' in \emph{Proc. Proc.
  IEEE Inter. Conf. Commun. Workshops (ICC Workshops )}, 2019, pp. 1--6.

\bibitem{schober2002velocity}
H.~Schober and F.~Jondral, ``Velocity estimation for {OFDM} based communication
  systems,'' in \emph{Proc. IEEE Veh. Technol. Conf.}, vol.~2.\hskip 1em plus
  0.5em minus 0.4em\relax IEEE, 2002, pp. 715--718.

\bibitem{de2021joint}
L.~G. de~Oliveira, B.~Nuss, M.~B. Alabd, A.~Diewald, M.~Pauli, and T.~Zwick,
  ``Joint radar-communication systems: Modulation schemes and system design,''
  \emph{IEEE Trans. Microw. Theory Tech.}, vol.~70, no.~3, pp. 1521--1551,
  2021.

\bibitem{gaudio2020effectiveness}
L.~Gaudio, M.~Kobayashi, G.~Caire, and G.~Colavolpe, ``On the effectiveness of
  {OTFS} for joint radar parameter estimation and communication,'' \emph{IEEE
  Trans. Wireless Commun.}, vol.~19, no.~9, pp. 5951--5965, 2020.

\bibitem{chen2019channel}
J.~Chen, Y.-C. Liang, H.~V. Cheng, and W.~Yu, ``Channel estimation for
  reconfigurable intelligent surface aided multi-user mmwave {MIMO} systems,''
  \emph{IEEE Trans. Wireless Commun.}, pp. 1--1, 2023, doi:
  10.1109/TWC.2023.3246264.

\bibitem{braun2014ofdm}
K.~M. Braun, ``{OFDM} radar algorithms in mobile communication networks,''
  Ph.D. dissertation, Karlsruhe, Karlsruher Institut f{\"u}r Technologie (KIT),
  Diss., 2014, 2014.

\bibitem{kay1993fundamentals}
S.~M. Kay, \emph{Fundamentals of statistical signal processing: estimation
  theory}.\hskip 1em plus 0.5em minus 0.4em\relax vol. 1. Englewood Cliffs, NJ,
  USA: Prentice-Hall., 1998.

\bibitem{ngo2013energy}
H.~Q. Ngo, E.~G. Larsson, and T.~L. Marzetta, ``Energy and spectral efficiency
  of very large multiuser {MIMO} systems,'' \emph{IEEE Trans. Commun.},
  vol.~61, no.~4, pp. 1436--1449, 2013.

\bibitem{grant2015cvx1}
M.~Grant, S.~Boyd, and Y.~Ye, ``{CVX}: {MATLAB} software for disciplined convex
  programming,'' \emph{[Online]. Available: http://stanford.edu/~boyd/cvx}.

\bibitem{chen2018resource}
J.~Chen, L.~Zhang, Y.-C. Liang, X.~Kang, and R.~Zhang, ``Resource allocation
  for wireless-powered {I}o{T} networks with short packet communication,''
  \emph{IEEE Trans. Wireless Commun.}, vol.~18, no.~2, pp. 1447--1461, 2019.

\bibitem{nguyen2017delay}
D.~H. Nguyen and R.~W. Heath, ``Delay and doppler processing for multi-target
  detection with {IEEE} 802.11 {OFDM} signaling,'' in \emph{Proc. Int. Conf.
  Acoust. Speech Signal Process. (ICASSP)}, 2017, pp. 3414--3418.

\bibitem{onubogu2014empirical}
O.~Onubogu, K.~Ziri-Castro, D.~Jayalath, K.~Ansari, and H.~Suzuki, ``Empirical
  vehicle-to-vehicle pathloss modeling in highway, suburban and urban
  environments at 5.8 ghz,'' in \emph{Proc. Int. Conf. Signal Process. Commun.
  Syst. (ICSPCS)}.\hskip 1em plus 0.5em minus 0.4em\relax IEEE, 2014, pp. 1--6.

\bibitem{godrich2011power}
H.~Godrich, A.~P. Petropulu, and H.~V. Poor, ``Power allocation strategies for
  target localization in distributed multiple-radar architectures,'' \emph{IEEE
  Trans. Signal Process.}, vol.~59, no.~7, pp. 3226--3240, 2011.

\end{thebibliography}
\bibliographystyle{IEEEtran}%By using IEEEtrans, the number can be displayed.

\end{document}